\documentclass[a4paper,10pt,notitlepage]{article}
\usepackage[a4paper, left=3.5cm, right=3.5cm, top=2cm]{geometry}
\usepackage{amsfonts,amssymb,amsmath,amsthm,hyperref,colonequals}
\usepackage{caption}
\usepackage{cite}
\usepackage{color}
\usepackage{tabularx}
\usepackage{graphicx}
\usepackage{bbm}
\usepackage{cleveref}
\usepackage[export]{adjustbox}
\usepackage{shuffle}
\usepackage[nice]{nicefrac} 
\usepackage{url}
\usepackage[normalem]{ulem}
\usepackage{mathtools}
\theoremstyle{definition}
\newcommand{\bea}{\begin{eqnarray}}
\newcommand{\eea}{\end{eqnarray}}
\newcommand{\be}{\begin{equation}}
\newcommand{\ee}{\end{equation}}

\DeclareMathOperator{\Tr}{Tr}

\catcode`,\active

\catcode`\,12

\begin{document}

\thispagestyle{empty}
\setcounter{page}{0}
\begin{flushright}\footnotesize
\vspace{0.5cm}
\end{flushright}
\setcounter{footnote}{0}

\begin{center}
{\LARGE{
\textbf{Field-theoretic approach to large-scale structure formation} 
}}

\bigskip
\medskip

{\sc \large
Pavel Friedrich$^*$ and Tomislav Prokopec$^\ddagger$ }\\[5mm]

{\it Institute for Theoretical Physics, Spinoza Institute and the
Center for Extreme Matter and Emergent Phenomena (EMME$\Phi$),\\ 
Utrecht University, Buys Ballot Building,
Princetonplein 5, 3584 CC Utrecht, The Netherlands
}\\[5mm]
\let\thefootnote\relax\footnotetext{\\$^*$Electronic address: \texttt{p.friedrich@uu.nl}\\
$^\ddagger$Electronic address: \texttt{t.prokopec@uu.nl}
}

\end{center}
\begin{abstract}
We develop a field-theoretic description of large-scale structure formation by taking the non-relativistic limit of a canonically transformed, real scalar field which is minimally coupled to scalar gravitational perturbations in longitudinal gauge. We integrate out the gravitational constraint fields and arrive at a non-local action which is only specified in terms of the dynamical degrees of freedom. In order to make this framework closer to the classical particle description, we construct the corresponding 2PI effective action truncated at two loop order for a non-squeezed state without field expectation values. We contrast the dynamical description of the coincident time phase-space density to the standard Vlasov description of cold dark matter particles and identify momentum and time scales at which linear perturbation theory will deviate from the standard evolution. 
\end{abstract}
 

\newpage
\setcounter{page}{1}


\tableofcontents
\addtolength{\baselineskip}{5pt}

\section{Introduction}
It lies in the nature of physics that surprising effects happen on the transition between one physical scale to another.
In order to study whether such transitioning effects are important one ought to start from the most fundamental description that is available and descend in a controlled way to the scale that is relevant for the problem.
Cosmological theories are in particular sensible to such transitions since they attempt to describe various scales and its associated effects range from quantum field physics during inflation up to the evolution of large-scale structures and cold dark matter at later times which is what we are interested in. Even if one assumes only a real scalar particle with  gravitational interactions in  a non-relativistic limit, there is still room to choose the state which should describe this cold dark matter, be it a  classical stochastic state with or without squeezing, or a condensate. In \cite{Prokopec:2017ldn,Friedrich:2018qjv}, we showed that a non-squeezed, classical stochastic state leads to point-like cold dark matter characteristics on large scales and is thus the field-theoretic generalization of the standard Vlasov description \cite{Bernardeau:2001qr, lrr-2011-4}. The condensate description corresponding to a coherent state, on the other hand, is referred to as fuzzy dark matter \cite{Turner:1983he, Sin:1992bg, Lee:1995af, Hu:2000ke, Goodman:2000tg, Peebles:2000yy}. It also resembles point-like cold dark matter dynamics on large scales but there are, however, significant small scale effects \cite{Marsh:2015daa,MARSH20161, Hui:2016ltb, Marsh:2018zyw}. Are such small scale effects an exclusive features of a condensate state, do they occur for other states, how do they differ?

In order to account for these questions we are after a field-theoretic description of cold dark matter that originates from the QFT tree-level action of a real scalar field with minimal coupling to gravity where we focus on scalar gravitational perturbations in longitudinal gauge in an FLRW universe. We would like to emphasize that using an action of genuine quantum nature does not imply that quantum effects are considered important, field-theoretic effects, however, may be and we will give examples of such effects in this paper. One of the key ingredients in this work is the generalization of the canonical field transformation developed in \cite{Namjoo:2017nia} where the non-relativistic limit of a self-interacting real scalar field in Minkowski space-time is addressed. We perturb the general relativistic theory \eqref{fundamentalTheory} and rewrite it in terms of the diagonal field representation \eqref{defPsi}. We then take the non-relativistic limit assuming that the mass $m$ of the scalar is the largest scale  apart from the Planck scale $M_P$. The resulting action \eqref{finalPsiGenState} contains the classical, non-relativistic particle description as a special case on large-scales. We show this by constructing the corresponding 2PI effective action truncated at two loop order for a virialized state, namely a state that is neither squeezed nor that it has a non-vanishing condensate. Viriliazed states can contain a large number of particles, if they descend from a mixed density matrix.
 
The work we present in this paper is in line with our previous works \cite{Prokopec:2017ldn,Friedrich:2018qjv}. However, the main differences are first, that we perturbatively integrate out the gravitational constraint fields which leads to an additional exchange interaction and second, that we set up a general framework where we {\it  a priori} do not assume that spatial gradients $\nabla_{\vec{X}}$ are small compared to the particle momenta $\vec{p}$ which is important if one would like to study small scale effects. 

Let us also mention that the development of the framework in this paper is also motivated by the problem of solving cold dark matter dynamics beyond the linearized, single-stream perfect fluid approximation. Similar to the statistical field theory based on classical point-like particles \cite{Geiss:2018spv,Bartelmann:2019unp} and as an extended approach to the condensate based Schroedinger model \cite{Uhlemann:2014npa,Garny:2017xkc,Uhlemann:2018olp,Uhlemann:2018gzz}, we reformulate the problem of cold dark matter dynamics by resorting to a more fundamental description which may be more suitable to get a different analytical and numerical access. 

We work in units where $c=1$ with a mostly plus signature $(-,+,+,+)$.
\pagebreak

\section{Gravity through external fields }
Let us start by writing down the action for a massive, real scalar field in its canonical form with couplings to gravity in ADM-variables \cite{Arnowitt:1962hi},
\be
S_{\phi} = \int^{t_2}_{t_1} dt \int_{\Sigma_t} d^{3} x  \Big[  \Pi_{\phi} \dot{\phi} - \frac{N}{2}{\gamma^{1/2}} \Big({\gamma^{-1}} \Pi_{\phi}^2
+  \gamma^{ij} \partial_i \phi 
\partial_j \phi +  \frac{m^2}{\hbar^2} \phi^2\Big)-N_i \Pi_{\phi} \partial^i \phi \Big] \, , \label{fundamentalTheory}
\ee
where $N$ and $N^i$ are lapse and shift functions, $\gamma_{ij}$ is the spatial metric, $\gamma$ its determinant  and $\pi_{\phi}$ is the canonical momentum associated with $\phi$.
We now neglect vector and tensor perturbations in the metric and consider scalar perturbations in the longitudinal gauge   with the gravitational potentials $\Phi_G$ and $\Psi_G$, in which we also linearize with a small perturbation parameter $\varepsilon_g$,
\be
  N = \overline{N}(1 + \Phi_G) \,, \quad N^i = 0 \, , \quad
\gamma_{ij} = a^2  \delta_{ij}(1 - 2 \Psi_G) \, , \quad
 {\gamma}^{1/2}= a^{3} \big[ 1 - 3 \Psi_G  \big] \, , \label{scalarPerts}
\ee
\be
\mathcal{O} \big( \Phi_G\, , \Psi_G \big) = \varepsilon_g^2 \ll 1\, .
\ee
This leads us to
\begin{multline}
S_{\phi} \approx \int^{t_2}_{t_1} dt \int_{\Sigma_t} d^{3} x  \Big[  \Pi_{\phi} \dot{\phi} - \frac{1}{2}\overline{N}(1 + \Phi_G) \Big( a^{-3}\big[ 1 + 3 \Psi_G  \big] \Pi_{\phi}^2 \\
+ a \big[ 1 -  \Psi_G  \big]\delta^{ij} \partial_i \phi 
\partial_j \phi +  a^{3}\big[ 1 - 3 \Psi_G  \big]\frac{m^2}{\hbar^2} \phi^2\Big) \Big] \, .
\end{multline}
We switch to conformal time $a d \eta = \overline{N} dt$ whose derivative is denoted by a prime ($a \mathcal{H} = a^{\prime}$)
and perform a first canonical transformation (leaving the path-integral measure unchanged) by defining
\be
\phi_c \equiv a \phi\,, \quad \Pi_{\phi}^c \equiv a^{-1} \Pi_{\phi} + \mathcal{H} a\phi \, .
\ee
We integrate by parts and find upon dropping temporal boundary terms
\begin{multline}
S_{\phi} \approx S_{\phi_c} \equiv  \int^{{\eta}_2}_{{\eta}_1} d{\eta} \int_{\Sigma_{\eta}} d^{3} x  \Big[ \Pi_{\phi}^c\phi_c^{\prime}  - \frac{1}{2}  \Big\lbrace \big[1+ \Phi_G + 3 \Psi_G  \big] \big(\Pi_{\phi}^c \big)^2  - 2 \big[\Phi_G + 3 \Psi_G  \big]\mathcal{H}   \phi_c \Pi_{\phi}^c  \\
+ \big[ 1+ \Phi_G -  \Psi_G  \big]\delta^{ij}  \partial_i \phi_c 
\partial_j \phi_c +  \big[ 1 +\Phi_G - 3 \Psi_G  \big]\frac{m^2_{\text{eff}} }{\hbar^2} \phi_c^2 +  \big[ \big(   \mathcal{H}^{\prime}+ 2\mathcal{H}^2 \big) \Phi_G  -3  \mathcal{H}^{\prime} \Psi_G \big] \phi_c^2   \Big\rbrace   \Big] \, , \label{Sphic}
\end{multline}
where we identify the effective mass 
\be
m^{2}_{\text{eff}} \equiv m^2 a^2 - \hbar^2 \mathcal{H}^{\prime} - \hbar^2 \mathcal{H}^2\, .
\ee
We now propose a straightforward generalization of the non-local field redefinition worked out for Minkowski space-time by \cite{Namjoo:2017nia},
\be
\psi \equiv \frac{ 1}{\sqrt{ 2 \hbar }}{\mathcal{E}}^*
 \hat{\Omega}^{1/2} \Big( \phi_c + i {\hbar} \hat{\Omega}^{-1} \Pi^c_{\phi} \Big)\, , \quad \hat{\Omega}  \equiv \sqrt{ m_{\text{eff}}^2 - \hbar^2 \Delta}\, , \label{defPsi}
\ee
where the spatial Laplacians is given by
\be
\Delta \equiv \delta^{ij} \partial_i \partial_j\, , 
\ee
and the time-dependent phase $\mathcal{E} $ is defined as
\be
\mathcal{E}(\eta) \equiv \exp \Big(-i \int^{\eta} \frac{m_{\text{eff} }(\widetilde{\eta})}{\hbar}d \widetilde{\eta} \Big)\,.
\ee
The transformation \eqref{defPsi} is akin to going to creation and annihilation operator variables in which one may diagonalize the Hamiltonian in the free theory. Moreover, it removes {\it Zitterbewegung} generated by the mass term. The operator $\hat{\Omega}$ has the interpretation of a particle energy.
The reverse transformation of \eqref{defPsi} reads
\be
\phi_c =  \sqrt{\frac{\hbar}{2 \hat{\Omega} }} \Big({\mathcal{E}} \psi + {\mathcal{E}^*} \psi^{*}    \Big) \,, \quad 
\Pi^c_{\phi} = - i \sqrt{\frac{ \hat{\Omega}}{2\hbar }} \Big({\mathcal{E}} \psi - {\mathcal{E}^*} \psi^{*}    \Big)  \,.
\ee
We note, that the corresponding measure in the path-integral is in the Hamilton formulation related to the real and imaginary parts of $\psi$,
\be
\mathcal{D} \phi_c \mathcal{D} \Pi_{\phi}^c \propto \mathcal{D} \text{Re} \psi \, \mathcal{D} \text{Im} \psi\, .
\ee
Thus, we have a canonical transformation between the fields $\Phi_c\, , \Pi_{\phi}^c$ and $ \text{Re} \psi \, ,\text{Im} \psi$.
Moreover, one obtains the expected, equal-time commutation relation for the corresponding quantum operators in the non-relativistic theory,
\be
\big[\hat{\psi}(\eta, x^i)\, ,\hat{\psi}^{\dagger}(\eta, {y}^i)  \big]  = \hbar  \delta^{3} \big(  x^i\, ,y^i \big)\, .
\ee
Plugging in the transformation \eqref{defPsi}  into the action \eqref{Sphic}, we find
\begin{multline}
S_{\phi_c} = S_{\psi}\big[\Phi_G, \Psi_G \big]  \equiv  
\int^{{\eta}_2}_{{\eta}_1} d{\eta} \int_{\Sigma_{\eta}} d^{3} x  \Bigg\lbrace i  \psi^{*} \psi^{\prime} - \frac{ m_{\text{eff}}}{\hbar} \psi \Big(\frac{\hat{\Omega}}{m_{\text{eff}}}-1 \Big)\psi^{*}    -  \frac{1}{2}\frac{ m_{\text{eff}}^{\prime}}{ m_{\text{eff}}} \text{Im} \, \Big[ {\mathcal{E}}^2 \psi \frac{m_{\text{eff}}^2}{\hat{\Omega}^2}\psi \Big] \\ - \frac{m_{\text{eff}}}{ \hbar}  \Big[\big[ \Phi_G + 3 \Psi_G  \big] \Big[\sqrt{\frac{\hat{\Omega}}{m_{\text{eff}}}}\text{Im} \big({\mathcal{E}} \psi   \big) \Big]^2  +  \big[\Phi_G - 3 \Psi_G  \big]\Big(\sqrt{\frac{m_{\text{eff}}}{\hat{\Omega}} }\text{Re} \big({\mathcal{E}} \psi   \big) \Big)^2\Big]  \\
-    \frac{\hbar}{ m_{\text{eff}}} \big[ \big(   \mathcal{H}^{\prime}+ 2\mathcal{H}^2 \big) \Phi_G  -3  \mathcal{H}^{\prime} \Psi_G \big] \Big[\sqrt{\frac{m_{\text{eff}}}{\hat{\Omega}}} \text{Re}\big({\mathcal{E}} \psi   \big) \Big]^2 
- \frac{\hbar}{m_{\text{eff}}}(\Phi_G-\Psi_G) \delta^{ij} \sqrt{\frac{m_{\text{eff}}}{\hat{\Omega}}} \partial_i  \text{Re} \big({\mathcal{E}} \psi  \big)\sqrt{\frac{m_{\text{eff}}}{\hat{\Omega}}}\partial_j   \text{Re} \big({\mathcal{E}} \psi  \big) \\+ 2 \big[\Phi_G + 3 \Psi_G  \big]  \mathcal{H}  \Big[\sqrt{\frac{m_{\text{eff}}}{\hat{\Omega}}} \text{Re}\big({\mathcal{E}} \psi   \big) \Big]\Big[\sqrt{\frac{\hat{\Omega}} {m_{\text{eff}}}}\text{Im}\big({\mathcal{E}} \psi   \big) \Big]  \Bigg\rbrace\, .
\end{multline}
The transformation \eqref{defPsi} was designed to obtain a non-relativistic description in $\hbar^2 \|\Delta\| \ll m_{\text{eff}}^2 $ such that one can perturbatively correct it in a controlled way.
Spatial derivatives $\nabla = \nabla_{\vec{x}}$ acting on matter fields $\psi(\vec{x})$ will be mapped on particle momenta $\vec{p}$ and long-distance gradients $\nabla_{\vec{X}}\sim \hbar \vec{k}$ once two-point functions of fields such as $\langle {\psi}^{\dagger} (\eta,\vec{x})  {\psi}(\eta,\vec{y}))\rangle$ are mapped to a particle phase-space density $f(\eta, \vec{p},\vec{X})$. Thus, assuming $\hbar^2 \|\Delta\| \ll m_{\text{eff}}^2 $ corresponds to assuming physical momenta $p$ and inverse distance scales $L^{-1}\sim k$ of the underlying physical problem to be much smaller than the scale set by the mass $m_{\text{eff}}$.

Let us subsume these scale relations in the following expansion parameter
\be
\mathcal{O} \Big(  \frac{\hbar\|\nabla\|}{m} \Big) = \varepsilon_{\text{nr}} \ll 1\, .
\ee 
We will only keep leading order contributions in $\varepsilon_{\text{nr}}$  and also drop multiplicative higher-order terms of the type $\varepsilon_g^2 \cdot \varepsilon_{\text{nr}}^2$  that involve the gravitational perturbation parameter.
Moreover, we want to consider the case where the mass $m$ is much bigger than the Hubble rate or its logarithmic derivative 
\be
\mathcal{O} \Big(   \frac{\hbar\mathcal{H}}{m a}\, , \frac{\hbar\mathcal{H}^{\prime}}{\mathcal{H} m a }\Big) = \varepsilon_{\text{\scriptsize H/m}} \ll 1\, ,
\ee
In what follows, we shall  keep only leading order contributions of order $ \varepsilon_{H/m}$ and drop  multiplicative higher-order terms of order $\varepsilon_g^2 \cdot \varepsilon_{\text{\scriptsize H/m}}^2$ involving the gravitational potential. However, we keep terms of order $\varepsilon_g^2 \cdot \varepsilon_{\text{\scriptsize H/m}}$ since they come with phase-factors whose time derivative can reduce the order by one power.
We then have
\begin{multline}
S_{\psi}\big[\Phi_G, \Psi_G \big] \approx\int^{{\eta}_2}_{{\eta}_1} d{\eta} \int_{\Sigma_{\eta}} d^{3} x  \Bigg\lbrace i  \psi^{*} \psi^{\prime} +  \psi^{*}  \Big(\frac{ \hbar \Delta}{2 m a}- \frac{m a}{\hbar} \Phi_G \Big)\psi     + 3 \frac{m a }{ \hbar}  \Psi_G \text{Re} \, \big( {\mathcal{E}}^2 \psi^2  \big) \\
-\mathcal{H}    \Big(\frac{1}{2} -\Phi_G - 3 \Psi_G  \Big) \text{Im} \, \big( {\mathcal{E}}^2 \psi^2  \big) \Bigg\rbrace\, . \label{sPsiWithPot}
\end{multline}
What we have achieved so far is a different viewpoint on the non-relativistic limits we discussed in \cite{Prokopec:2017ldn}  and \cite{Friedrich:2018qjv} by assuming small gradients and a small expansion rate of scale factor with respect to the mass.
If we promote the field $\psi$ to an operator, we find that we treated the equal-time correlators 
\be
\langle \hat{\Pi}_{\phi} (x)   \hat{\Pi}_{\phi} (y) \rangle \,, \;  \langle \hat{\Pi}_{\phi} (x)  \hat{\phi} (y) \rangle \,, \;  \langle \hat{\phi} (x)  \hat{\Pi}_{\phi} (y) \rangle\,, \;  \langle \hat{\phi} (x)  \hat{\phi} (y) \rangle\ , 
\ee
for the equal-time correlators
\be
\langle \hat{\psi} (x) \hat{ \psi}^{\dagger}  (y) \rangle \,, \;  \langle \hat{\psi}^{\dagger}  (x)  \hat{\psi} (y) \rangle \,, \;  \langle \hat{\psi} (x)  \hat{\psi} (y) \rangle\,, \;  \langle \hat{\psi}^{\dagger}  (x)  \hat{\psi}^{\dagger}  (y) \rangle\, . \label{allcorrs}
\ee
In \cite{Prokopec:2017ldn,Friedrich:2018qjv}  we concluded that only a particular combination of suitably transformed correlators constitutes a phase-space density of classical particles, the other ones being highly oscillatory and suppressed if they are initially small. The situation is similar in the new variables and amounts to neglecting $\langle \hat{\psi} (x)  \hat{\psi} (y) \rangle$ and $\langle \hat{\psi}^{\dagger}  (x)  \hat{\psi}^{\dagger}  (y) \rangle$ in comparison to $\langle \hat{\psi}  (x) \hat{\psi}^{\dagger}  (y) \rangle$ and $\langle \hat{\psi}^{\dagger}  (x) \hat{\psi}  (y) \rangle$. It is usually the case that if one drops these squeezing contribution, one can show that if they are not present initially, the evolutions will generate them only under special circumstances. Apart from the limits we have taken so far, we can consider this requirement on the quantum state as another requirement to obtain a description of classical particles from a real scalar quantum field. We refer to such a state as a {\it virialized} state since the kinetic energy in field space expressed through the $ {\Pi}_{\phi}{\Pi}_{\phi}$-correlator is of the same order as the potential energy expressed through particle energy squared times the ${\phi}{\phi}$-correlator. A virialized state corresponds to a spherical blob in the phase-space diagram of the real scalar field. This state is more general than a thermal state since no relationship is assumed between phase-space occupancy of different field momenta. Thus, assuming the oscillatory correlators to be small initially, we can omit them from the dynamical description,
\be
S_{\psi} \big[\Phi_G, \Psi_G \big] \stackrel{\text{virialized state}}{\approx} \int^{{\eta}_2}_{{\eta}_1} d{\eta} \int_{\Sigma_{\eta}} d^{3} x  \Big[ i  \psi^{*} \psi^{\prime} +  \psi^{*}  \Big(\frac{ \hbar \Delta}{2 m a}- \frac{m a}{\hbar} \Phi_G \Big)\psi   \Big]\, ,
\ee
and the operator equation corresponding to this action reads (for classical gravitational fields),
\be
i \partial_{\eta} \hat{\psi} (\eta, x^i ) = - \Big[\frac{\hbar \Delta_x}{2 m a}- \frac{m a}{\hbar} \Phi_G (\eta, x^i ) \Big] \hat{\psi}(\eta, x^i )\, .
\ee
Choosing a coherent quantum state such that the connected piece of the two-point functions are negligible and classical fields are a good enough approximation leaves us with the dark matter description   coined {\it fuzzy dark matter}.
However, as we advocated in \cite{Prokopec:2017ldn,Friedrich:2018qjv}, we do not have to restrict ourself to one-point functions since choosing a more-general state allows {\it{a priori}} for vorticity and anisotropy without additional course graining.
For such a more general state with non-vanishing connected two-point functions, we can define a Wigner transformation (which corresponds to the spatially covariant one in \cite{Friedrich:2018qjv} to zeroth order in gravitational perturbations),
\be
f(\eta, X^i, p_i) \equiv  \frac{1}{(2 \pi \hbar)^3 \hbar}\int d^3 r e^{- \frac{i}{\hbar} r^k p_k} \langle {: \widehat{\psi} (\eta, X^i + {r^i/2} ) \, {\widehat{\psi}}^{\dagger} (\eta, X^i - {r^i/2} ) :}\rangle\, ,
\ee
where we made use of a local normal ordering prescription "$::$" that essentially subtracts the state-independent quantum contribution of the two-point function such that a gradient expansion in $\hbar p_i \partial_{X^i}$ is possible (in other words, we have a hierarchy of scales $ma \gg p \gg \hbar \partial_X$ together with $ma \gg \mathcal{H}$, for more details see   \cite{Friedrich:2018qjv}).
The dynamical equation for the phase-space density $f$ approaches the Vlasov equation for cold dark matter to leading order in the spatial gradient expansion
\be
\Bigg[ \frac{\partial}{\partial \eta} 
+ \frac{p_k}{m a } \frac{\partial}{\partial X^k}-m a \big[1+ \mathcal{O}\big(\hbar^2 \big) \big] \frac{\partial}{\partial X^k} \Phi_G (\eta, X^i) \frac{\partial}{\partial p_k} \Bigg]f(\eta, X^i, p_i) = 0\, . \label{Vlasov}
\ee
\pagebreak
\section{Integrating out gravitational fields \label{genStateSecACtionc}}
Instead of treating the gravitational perturbations as part of a classical (possibly stochastic) background metric, we treat them now as quantum fluctuations and integrate them out. This approach enables one to be more accurate in comparison to the one-loop semi-classical expansion and leaves only the true degrees of freedom in the description of the theory. 
The starting point for the gravitational part is the Einstein-Hilbert action in the ADM formulation 
\be
S_{g} = \int^{t_2}_{t_1} dt \int_{\Sigma_t} d^{3} x  \Big[ \Pi^{ij} \dot{\gamma}_{ij}  - N \mathcal{H}_0^{(g)}- N^i \mathcal{H}_i^{(g)} \Big] + \int^{t_2}_{t_1} dt \int_{\partial \Sigma_t} d^{2} x \mathcal{H}_B \, , \label{gravAction}
\ee
where the spatial  boundary term  $\mathcal{H}_B$  specified in \cite{Dyer:2008hb} is of no relevance for us and the Hamilton and momentum constraints  of the gravitational sector are given by
\bea
\mathcal{H}_0^{(g)} &=&- \frac{M_P^2}{2 \hbar}  \gamma^{1/2}{R^{(n-1)}} 
+ \frac{2\hbar }{M_P^2 \gamma^{1/2}} \big[  \Pi_{ij} \Pi^{ij} - \frac{ \Pi^2}{2} \big] \,  , \\
\mathcal{H}_i^{(g)} &=&  -2 \gamma^{1/2} {^{(3)} \nabla^j} \frac{\Pi_{ij} }{\gamma^{1/2}}\,, \label{h0hi}
\eea
which should not be confused with the conformal Hubble rate $\mathcal{H}$. In the gravitational Hamiltonian densities \eqref{h0hi}, we made use of the reduced Planck mass $M_P$  and the canonical momentum $\Pi^{ij}$ conjugate to the spatial metric $\gamma_{ij}$.  We also denoted the trace of the canonical momentum as $\Pi = \gamma_{ij}\Pi^{ij}$ and introduced the covariant derivative ${^{(3)} \nabla }$ on spatial sections.
As a first step to a non-relativistic limit of gravitating matter in an expanding universe, we will approximate the gravitational action \eqref{gravAction} as in the semi-classical case with scalar perturbations in the longitudinal gauge. In addition to the decomposition of lapse, shift and spatial metric in \eqref{scalarPerts}, we also need to compose the canonical momentum of the spatial metric which we do as follows,
\be
\Pi^{ij} = \delta^{ij} a^{-2} \Pi_a \Big(1+ \frac{1}{2}\Psi_G +\frac{1}{2}\Pi_{\Psi} \Big)\, .
\ee
A few comments on this split into a homogeneous background $\overline{N}, a , \Pi_a$ and the path-integral perturbations $\Phi_G, \Psi_G, \Pi_{\Psi}$ are in order.
 The    obvious difference to the semi-classical analysis lies in the fact that we are treating inhomogeneous perturbations not any more as part of the classical (external) background which allows one to go beyond semi-classical one-loop approximation and include {\it{in principle}} quantum effects. This, however, does {\it{not}} mean that these perturbations {\it{necessarily }}correspond  to quantum-sized effects. Whether such effects are important depends  on the initial conditions: so are vacuum fluctuations the essential ingredient for inflationary models, whereas they are in most scenarios not at all for non-relativistic set-ups with a highly populated state ("many particles"). 
 Let us also mention some boundary conditions of the perturbations  $\Phi_G, \Psi_G, \Pi_{\Psi}$. We will assume that a well chosen background will keep any  zero-mode fluctuations negligible such that the perturbations  $\Phi_G, \Psi_G, \Pi_{\Psi}$ decay at spatial infinity at least as $1/r$. For the same reason we will ignore the boundary term in \eqref{gravAction}. Having said this, we will already make a choice for  the background field $\Pi_a$  such that it evolves according to the background equations of motion
\be
\Pi_a = - \frac{M_P^2}{ \hbar} a^2 \mathcal{H}\, .
\ee
After these remarks we expand the gravitational action \eqref{gravAction} in conformal time for longitudinal scalar perturbations up to quadratic order (cf. \cite{Anderegg:1994xq,Armendariz-Picon:2016dgd}), drop the zero order contribution $\bar{S}_g$ from the gravitational part  and add the matter action \eqref{sPsiWithPot}, 
\begin{multline}
S\Big[\Phi_G, \Psi_G,\Pi_{\Psi},\psi \Big] \equiv S_{\psi}\Big[\Phi_G, \Psi_G,\psi \Big]+ S_g \Big[\Phi_G, \Psi_G, \Pi_{\Psi} \Big] -\bar{S}_g   \\ \approx    \int^{{\eta}_2}_{{\eta}_1} d{\eta} \int_{\Sigma_{\eta}} d^{3} x  \Bigg\lbrace i  \psi^{*} \psi^{\prime} +  \psi^{*}  \Big(\frac{ \hbar \Delta}{2 m a}- \frac{m   a}{\hbar} \Phi_G \Big)\psi  + 3 \frac{m a }{ \hbar}  \Psi_G \text{Re} \, \big( {\mathcal{E}}^2 \psi^2  \big) \\
  - \mathcal{H}    \Big(\frac{1}{2}-\Phi_G - 3 \Psi_G  \Big) \text{Im} \, \big( {\mathcal{E}}^2 \psi^2  \big)  \Bigg\rbrace\\
  + \frac{M_P^2}{ 2\hbar}
\int^{{\eta}_2}_{{\eta}_1} d{\eta} \int_{\Sigma_{\eta}} d^{3} x  \Bigg\lbrace - 6a^2  \big( \mathcal{H}^2 +2  \mathcal{H}^{\prime}\big)  \Psi_G 
+ 6a^2 \mathcal{H} (\Pi_{\Psi}+ \Psi_G) \big(  \mathcal{H}\Psi_G  + \Psi_G^{\prime}  \big) 
+ \frac{3}{2} a^{2}\mathcal{H}^2(\Pi_{\Psi}+ \Psi_G)^2  \\
+ 6  a^{2}\mathcal{H}^2 \Phi_G(1 +   \Pi_{\Psi} )   
-2a^{2} 
    \Psi_G \Delta \Psi_G  - 3 a^{2}\mathcal{H}^2\Psi_G^2 
+  4  a^{2}  \Phi_G 
 \Delta \Psi_G\Bigg\rbrace\, .\label{fullACtion}
\end{multline}
We make the important remark that we did not expand the matter field $\psi$ around a background value. The main reason why we do this lies in the observation that the  perturbative expansion in \eqref{fullACtion} is valid if we supply the matter fields with appropriate boundary which are more general than a spatially homogeneous expectation value.  We will shortly come back to this issue.

If we now vary with respect to $\Phi_G$, we get the following constraint  
\be
   a^{2}  \Delta \Psi_G + \frac{3}{2}  a^{2}\mathcal{H}^2 (1+ \Pi_{\Psi} )- \frac{ \hbar}{2 M_P^2}\frac{m   a}{\hbar}\psi^{*}  \psi + \frac{ \hbar}{2 M_P^2}\mathcal{H}     \text{Im} \, \big( {\mathcal{E}}^2 \psi^2  \big) = 0\, , \label{constraint}
\ee
which means at the level of path integrals, that we generate a delta function by integrating over $\Phi_G$.
We have
\be   \Pi_{\Psi} = \frac{E_0 (\psi)}{3 a \mathcal{H}^2} - \frac{2}{3}\frac{{\Delta} \Psi_G
 }{ \mathcal{H}^2}  \, , \label{constraintRev}
\ee
where we defined
\be
E_0 (\psi) \equiv \frac{ \hbar}{M_P^2}\frac{m   }{\hbar}\psi^{*}  \psi -3  a \mathcal{H}^2 - \frac{\mathcal{H} }{a} \frac{ \hbar}{M_P^2}    \text{Im} \, \big( {\mathcal{E}}^2 \psi^2  \big)\,.
\ee
Let us also define
\be
E_1(\psi) \equiv a \mathcal{H}^2 +2 a \mathcal{H}^{\prime}- \frac{m }{\hbar} \frac{ \hbar}{M_P^2}\text{Re} \, \big( {\mathcal{E}}^2 \psi^2  \big)  - \frac{\mathcal{H}}{a}   \frac{ \hbar}{M_P^2}   \text{Im} \, \big( {\mathcal{E}}^2 \psi^2  \big) \,.
\ee
We are now in the position to integrate out the gravitational fields $\Phi_G$ and 
$\Psi_G$ by plugging the constraint equation \eqref{constraintRev} back into the action \eqref{fullACtion},
\begin{multline}
S[\Phi_G, \Psi_G,\Pi_{\Psi},\psi ]\longrightarrow S[\Psi_G,\psi]  =    \int^{{\eta}_2}_{{\eta}_1} d{\eta} \int_{\Sigma_{\eta}} d^{3} x  \Big[ i  \psi^{*} \psi^{\prime} +  \psi^{*}  \frac{ \hbar \Delta}{2 m a}\psi  
  - \frac{1}{2} \mathcal{H}   \text{Im} \, \big( {\mathcal{E}}^2 \psi^2  \big)  \Big]\\
  + \frac{M_P^2}{ 2\hbar}
\int^{{\eta}_2}_{{\eta}_1} d{\eta} \int_{\Sigma_{\eta}} d^{3} x  \Big[ - 6 a  E_1(\psi) \Psi_G 
+ 6a^2 \mathcal{H} \big((3 a \mathcal{H}^2)^{-1}E_0 (\psi)- 2(3 \mathcal{H}^2)^{-1}{\Delta} \Psi_G  + \Psi_G\big) \big(  \mathcal{H}\Psi_G  + \Psi_G^{\prime}  \big) 
\\
+ \frac{3}{2} a^{2}\mathcal{H}^2\big((3 a \mathcal{H}^2)^{-1}E_0 (\psi)-2 (3 \mathcal{H}^2)^{-1}{\Delta} \Psi_G + \Psi_G\big)^2  
- 2 a^{2}
    \Psi_G \Delta \Psi_G  - 3 a^{2}\mathcal{H}^2\Psi_G^2 \Big]\, .
\end{multline}
Since both, $E_0$ and $E_1$ multiply terms linear in the gravitational perturbations, their homogeneous limit will be related to the Einstein equations as we will see shortly.
We simplify certain expressions and integrate by parts to make manifest that the gravitational potential  is an auxiliary field,
\begin{multline}
S[\Psi_G,\psi]  =    \int^{{\eta}_2}_{{\eta}_1} d{\eta} \int_{\Sigma_{\eta}} d^{3} x  \Big[ i  \psi^{*} \psi^{\prime} +  \psi^{*}  \frac{ \hbar \Delta}{2 m a}\psi  
  - \frac{1}{2} \mathcal{H}   \text{Im} \, \big( {\mathcal{E}}^2 \psi^2  \big)  \Big]\\
  + \frac{M_P^2}{ 2\hbar}
\int^{{\eta}_2}_{{\eta}_1} d{\eta} \int_{\Sigma_{\eta}} d^{3} x  \Big[
\frac{E_0^2(\psi)}{6 \mathcal{H}^2}
 - a \mathcal{H}^{-2} \Big[\frac{2}{3 }  \Delta E_0(\psi)+ 6  \mathcal{H}^{2} E_1(\psi)  -2 \mathcal{H}^{\prime}  E_0(\psi)+2     \mathcal{H}   E_0^{\prime}(\psi) \Big] \Psi_G 
 \\
+ \frac{2}{3 \mathcal{H}^2} a^2 \Psi_G \Delta^2 \Psi_G 
-2 a^2 \frac{\mathcal{H}^{\prime}}{\mathcal{H}^2} \Psi_G \Delta \Psi_G 
- 2 a^{2} 
    \Psi_G \Delta \Psi_G  -3 a^{2}\mathcal{H}^2\Psi_G^2  \Big]\, .\label{actionPsiPsiG}
\end{multline}
Varying the Hubble action \eqref{actionPsiPsiG} with respect to the gravitational potential $\Psi_G$ yields the following constraint equation,
\begin{multline}
\frac{2}{3 } a^2  {\Delta}^2 \Psi_G
-2 a^2 {\mathcal{H}^{\prime}}   \Delta \Psi_G 
-2 a^{2} 
   \mathcal{H}^2 \Delta \Psi_G  -3 a^{2}\mathcal{H}^4\Psi_G \\ 
  - a \Big[\frac{1}{3 }  \Delta E_0(\psi)+ 3  \mathcal{H}^{2} E_1(\psi)  - \mathcal{H}^{\prime}  E_0(\psi)+     \mathcal{H}   E_0^{\prime}(\psi) \Big]    =0 \, .\label{PoissonGen}
\end{multline}
If we want to integrate out the gravitational potential via the constraint equation \eqref{PoissonGen}, we have to invert the Laplace operator and assume that the quantities $ E_0(\psi)$, $ E_1(\psi)$ vanish at least as $1/r$ at spatial infinity since we made the same assumptions for the gravitational perturbations. In other words, we have to impose
\be
 E_0^{\infty} (\psi) \equiv \lim_{\|\vec{x}\| \rightarrow \infty} E_0 [\psi (\vec{x})] \stackrel{!}{=} 0\,, \label{boundaryCond1}
\ee
and
\be
 E_1^{\infty}(\psi) \equiv \lim_{\|\vec{x}\| \rightarrow \infty} E_1 [\psi (\vec{x})] \stackrel{!}{=} 0\,. \label{boundaryCond2}
\ee
We were implicitly always dealing with path integrals in this derivation and remark that the conditions \eqref{boundaryCond1} and \eqref{boundaryCond2} are in fact operator equations which involve more than the zero mode of the field $\psi$. Subtracting the gravitational background fields, we have\footnote{Note, that we decided to give here a simpler treatment than for example in \cite{Friedrich:2018qjv}, where we gave some remarks on the renormalization of coincident limit operator products in a similar set-up.} 
\bea
\hat{\rho}_{\infty} &\equiv& E_0^{\infty} (\hat{\psi}) +3  a\mathcal{H}^2 \nonumber \\&=& \frac{ \hbar}{M_P^2}   \int d^3 p \Big[ \frac{m   }{\hbar}:\hat{\psi}^{\dagger}(\vec{p}) \hat{\psi}(-\vec{p}) : - \frac{\mathcal{H}}{a}   \text{Im} \, \big( {\mathcal{E}}^2 :\hat{\psi}(\vec{p})\hat{\psi}(-\vec{p}) : \big)\Big]\,, \\
\hat{P}_{\infty} &\equiv& E_1^{\infty}(\hat{\psi})-a \mathcal{H}^2 -2 a \mathcal{H}^{\prime} \nonumber\\ &=& -  \frac{ \hbar}{M_P^2}\int d^3 p \Big[ \frac{m }{\hbar}\text{Re} \, \big( {\mathcal{E}}^2 : \hat{\psi}(\vec{p})\hat{\psi}(-\vec{p}) : \big)+ \frac{\mathcal{H}}{a}\text{Im} \, \big( {\mathcal{E}}^2 : \hat{\psi}(\vec{p})\hat{\psi}(-\vec{p}) : \big) \Big] \,.
\eea
Taking expectation value and inserting the conditions \eqref{boundaryCond1} and \eqref{boundaryCond2}, we recover the semi-classical Einstein equations at spatial infinity,
\bea
3  a \mathcal{H}^2 &=& \langle \hat{\rho}_{\infty} \rangle\, ,\label{delteE0} \\ 
-a \mathcal{H}^2 -2 a \mathcal{H}^{\prime} &=&\langle \hat{P}_{\infty}\rangle\, .\label{delteE1}
\eea
We realize  that the operators $\hat{\rho}_{\infty}$ and $\hat{P}_{\infty}$ should not fluctuate around their expectation values. Rigorously speaking, only if even by small amounts, they of course do. However, in a more rigorous treatment, we would also have to include zero-mode fluctuations in the gravitational sector which we assumed to negligible from the very beginning. This then resolves the apparent inconsistency.

We can conclude that the boundary conditions \eqref{boundaryCond1} and \eqref{boundaryCond2} can be met if we adjust the background metric (which is a priori free to choose) to satisfy equations  \eqref{delteE0} and \eqref{delteE1} which are determined by the two-point functions of the matter field $\psi$ at spatial infinity. With these adjustments, we are in the position to integrate out the gravitational potential $\Psi_G$ in the action \eqref{actionPsiPsiG} by completing the squares,
\begin{multline}
S[\Psi_G,\psi] \longrightarrow S[ \psi]   =    \int^{{\eta}_2}_{{\eta}_1} d{\eta} \int_{\Sigma_{\eta}} d^{3} x  \Big[i  \psi^{*} \psi^{\prime} +  \psi^{*}  \frac{ \hbar \Delta}{2 m a}\psi  
  - \frac{1}{2} \mathcal{H}    \text{Im} \, \big( {\mathcal{E}}^2 \psi^2  \big)   \Big]\\
  +\frac{M_P^2}{ 4\hbar}
\int^{{\eta}_2}_{{\eta}_1} d{\eta} \int_{\Sigma_{\eta}} d^{3} x  \Bigg\lbrace  \frac{E_0^2(\psi)}{3 \mathcal{H}^2}  - \frac{3}{4 \mathcal{H}^2}\Big[\frac{2}{3 }  \Delta E_0(\psi)+2 \mathcal{H}^{\prime}  E_0(\psi)+ 6  \mathcal{H}^{2} E_1(\psi) +2     \mathcal{H}^{2}  E_0(\psi) \Big]   \\
\times  \Delta^{-2}_{\mathcal{H}} \Big[\frac{2}{3 }  \Delta E_0(\psi)+2 \mathcal{H}^{\prime}  E_0(\psi)+ 6  \mathcal{H}^{2} E_1(\psi) +2     \mathcal{H}^{2}  E_0(\psi) \Big]     \Bigg\rbrace\, , \label{actionPsiAllHubble}
\end{multline}
where we introduced the operator
\be
 \Delta^2_{\mathcal{H}} \equiv  {\Delta}^2 
-3  
   (\mathcal{H}^2+{\mathcal{H}^{\prime}}) \Delta - 18 \mathcal{H}^4  \, .
\ee
While equation \eqref{actionPsiAllHubble} represents the sought-for action, for the purpose of this paper, and to make progress, we focus on the sub-Hubble limit of action \eqref{actionPsiAllHubble} and introduce another perturbation parameter
\be
\mathcal{O}\Bigg(\, \frac{\mathcal{H}^2}{\|\Delta\| }\, , \frac{\mathcal{H}^{\prime}}{\|\Delta\| } \,\Bigg) = { \varepsilon_{\text{\scriptsize H/k}}} \ll 1\,.
\ee
We have
\be
 \Delta^{-2}_{\mathcal{H}} = \Delta^{-2}\big[1  +3 (\mathcal{H}^2+\mathcal{H}^{\prime}) \Delta^{-1} + \mathcal{ O}\big(\varepsilon^2_{\text{\scriptsize H/k}} \big) \big] \, .
\ee
We assume that  the back reaction between super- and sub-Hubble modes is negligible and work to leading order in $\varepsilon_{\text{\scriptsize H/k}}$. Upon integration by parts we find
\begin{multline}
 S[\psi]  \approx  S_{\psi} \equiv    \int^{{\eta}_2}_{{\eta}_1} d{\eta} \int_{\Sigma_{\eta}} d^{3} x  \Big[i  \psi^{*} \psi^{\prime} +  \psi^{*}  \frac{ \hbar \Delta}{2 m a}\psi  
  - \frac{1}{2} \mathcal{H}    \text{Im} \, \big( {\mathcal{E}}^2 \psi^2  \big)    -\frac{M_P^2}{ 4\hbar}
 \Big(   E_0 (\psi) + 6 E_1 (\psi)  \Big) {\Delta}^{-1}  E_0 (\psi)   \Big] \, . 
\end{multline}
Before we plug in the concrete expressions for $E_0$ and $E_1$, let us  for convenience  rescale the fields as
\be
\psi \rightarrow \hbar^{1/2} \psi\, ,
\ee
such that the two-point function has the dimensions of a number density. We then define
\be
 \rho_0 \equiv 3 a \mathcal{H}^2 \frac{M_P^2}{\hbar m}   \, ,
\ee
and find
\begin{multline}
 S_{\psi} =
   \hbar \int^{{\eta}_2}_{{\eta}_1} d{\eta} \int_{\Sigma_{\eta}} d^3 x  \Big\lbrace i  \psi^{*} \psi^{\prime} +  \psi^{*}  \frac{ \hbar \Delta}{2 m a} \psi   -\frac{1}{2} \mathcal{H}     \text{Im} \, \big( {\mathcal{E}}^2 \psi^2  \big)  \\
   - 
 \frac{m^2}{4 M_P^2}\big[ \psi^{*}  \psi - \rho_0 - \frac{\hbar\mathcal{  H}}{m a}\text{Im} \, \big( {\mathcal{E}}^2 \psi^2\big)           \big] \Delta^{-1} \big[ \psi^{*}  \psi - \rho_0 - \frac{\hbar\mathcal{  H}}{m a}\text{Im} \, \big( {\mathcal{E}}^2 \psi^2\big)        \big] 
 \\ -
 \frac{m^2}{2M_P^2}\Big[     \frac{1}{\mathcal{H}}\frac{d \rho_0}{d \eta}-3  \text{Re} \, \big( {\mathcal{E}}^2 \psi^2\big) -3 \frac{\hbar\mathcal{  H}}{m a} \text{Im} \, \big( {\mathcal{E}}^2 \psi^2\big)   \Big]     \Delta^{-1} \big[ \psi^{*}  \psi - \rho_0 - \frac{\hbar\mathcal{  H}}{m a}\text{Im} \, \big( {\mathcal{E}}^2 \psi^2\big)        \big] 
\Big\rbrace\, . \label{finalPsiGenState}
\end{multline}
The action \eqref{finalPsiGenState} is one of the principal results of this work and it serves as the starting point for a more general discussion of scalar field cold dark matter since it   makes less assumptions about the underlying state, we only assumed that its momenta are mainly distributed in a non-relativistic but also sub-Hubble window after the background contributions at spatial infinity have been subtracted. Let us identify some future lines of research. By starting from \eqref{finalPsiGenState} one can approach the theory in the 2PI formulation which captures the dynamics and interplay of the various contributions to the state, namely: the condensate $\langle \psi \rangle$ ( "fuzzy cold dark matter"),  the two-point function $\langle \hat{\psi} \hat{\psi}^{\dagger} \rangle$ corresponding to a virialized state ("particle cold dark matter" plus field-theoretic corrections) and squeezed two-point functions $\langle \hat{\psi} \hat{\psi} \rangle$, $\langle \hat{\psi}^{\dagger}  \hat{\psi}^{\dagger} \rangle$. Assuming mostly fuzzy cold dark matter, one can study its back reaction on particle dark matter and vice versa. Moreover, a field-theoretic description of cold dark matter can also lead to new insights on how dark matter behaves on different scales and, due to this reformulation, hopefully even to new techniques on how to tackle non-linear evolution on large scales. 
\section{2PI formulation for a virialized state \label{2PIsection}}
 In order to make the relation between the field-theoretic and the particle picture more concrete, we will study for simplicity an non-squeezed state having no condensate which we call a virialized state. We postpone the more general case for the future. Since interaction terms couple the various state contributions, they cannot be consistently set to zero but they remain, however, small if we assume a large mass in comparison to the Hubble rate as one can see in \eqref{finalPsiGenState},
 \be
 \| \langle \hat{\psi} \hat{\psi}^{\dagger} \rangle \| \gg  \| \langle \hat{\psi} \hat{\psi} \rangle\| \, , \quad \|\langle \hat{\psi} \hat{\psi}^{\dagger} \rangle\| \gg \|\langle \hat{\psi}^{\dagger}  \hat{\psi}^{\dagger} \rangle\|   \, , \quad  \langle \psi \rangle \approx 0 \,  . \label{noSqueezenoVev}
 \ee
From the point of view of Lagrangians, non-vanishing condensates are natural when the scalar field couples linearly to external sources (an example being the axionic coupling to gauge theory), the two-point function framework without condensate is more natural when the scalar field couples quadratically to external sources (such as in the theory of scalar electrodynamics).
 First of all, we note that the equations of motion for the scale factors \eqref{delteE0} and \eqref{delteE1} reduce to 
 \be
\rho_0 \equiv 3 a \mathcal{H}^2 \frac{M_P^2}{\hbar m}  \approx   \int d^3 p \,  \langle : \hat{\psi}^{\dagger}(\vec{p}) \hat{\psi}(-\vec{p})   : \rangle\approx \text{const}\, . \label{initCondNoSqueez}
\ee
Thus, the scale factor has to evolve as in a matter dominated universe
\be
a(\eta) = a_I \frac{\eta^2}{\eta_I^2}\, ,
\ee
and we choose $a_I=1$. Moreover, it will be convenient to define
\be
\beta \equiv \frac{\hbar \eta_I^2}{2 m } = \frac{6 M_P^2}{ m^2 \rho_0} \,.
\ee
Using these relations, the approximation \eqref{noSqueezenoVev} and writing out the inverse Laplace operator, we find that the action \eqref{finalPsiGenState} reads,
\begin{multline}
 S_{\psi} \approx  \hbar  \int^{{\eta}_2}_{{\eta}_1} d{\eta} \int_{\Sigma_{\eta}} d^3 x  \Bigg\lbrace i \psi^{*}  \psi^{\prime} +  \frac{ \beta }{\eta^2}\psi^{*} \Delta\psi + \frac{ 3\rho_0}{8 \pi  \beta} \int d^3 y 
    \frac{\big[ \rho_0^{-1}\psi^{*}(\vec{x}) \psi(\vec{x})-1 \big]\big[ \rho_0^{-1}\psi^{*}(\vec{y}) \psi(\vec{y})-1 \big]}{\|\vec{x}-\vec{y}\|} \Bigg\rbrace\, ,
\end{multline}
where we for simplicity suppressed the $\eta$-dependence.
In the Schwinger-Keldysh formulation, we then have  the following effective action truncated at two loops with $M_P^{-2} \propto \beta^{-1}$ being the loop counting parameter of gravity,
\begin{multline}
\Gamma[iG_{ij}^{cd} ] = \hbar \int^{{\eta}_2}_{{\eta}_1} d{\eta} \int_{\Sigma_{\eta}} d^3 x\int^{{\eta}_2}_{{\eta}_1} d{\eta^{\prime}} \int_{\Sigma_{\eta^{\prime}}} d^3 y \sum_{c,d = \pm}c \mathcal{D}^{ij}_{cd} (\eta, \vec{x},\eta^{\prime},\vec{y}) i G^{dc}_{ji} (\eta, \vec{x}, \eta^{\prime},\vec{y}) - i \frac{\hbar}{2}\Tr \Big[ \log \big( i G^{cd}_{ij}   \big) \Big]\\
- i \hbar \sum_{c=\pm} \frac{1}{8}\int d^4x_1...d^4x_4 \Big[iG^{cc}_{12}(x_1,x_2)iG^{cc}_{12}(x_3,x_4)+iG^{cc}_{21}(x_1,x_2)iG^{cc}_{21}(x_3,x_4)\\
+2iG^{cc}_{12}(x_1,x_2)iG^{cc}_{21}(x_3,x_4)
+2iG^{cc}_{11}(x_1,x_3)iG^{cc}_{22}(x_2,x_4) 
\Big]\Big[V_{H}^c (x_1,...,x_4)+V_{E}^c (x_1,...,x_4)\Big]\, , \label{2piAction}
\end{multline}
where we defined the (formally divergent) derivative operator
\begin{multline}
\mathcal{D}^{ij}_{cd} \equiv \frac{\delta_{cd}}{2} \begin{bmatrix}
    0   & - i \partial_{\eta} +\frac{\beta\Delta_x}{\eta^2}\\
    i  \partial_{\eta}  +\frac{\beta\Delta_x}{\eta^2}  &    0   \end{bmatrix} \delta(\eta - \eta^{\prime} )\delta^3 (\vec{x}-\vec{y}) 
     \\ -   \frac{3}{2\beta}  \delta_{cd}\big[\Delta^{-1}_x  (1)\big]\begin{bmatrix}
   0   & 1  \\
   1  &   0   \end{bmatrix} \delta(\eta - \eta^{\prime} )\delta^3 (\vec{x}-\vec{y})\, , \label{operatorD}
\end{multline}
which acts on the four propagators
\bea
iG_{ij}^{++} (x,y) &\equiv&\begin{bmatrix}
    \langle T\big[ \hat{\psi}( x)\,   \hat{\psi}(y)\big] \rangle    & \langle  T\big[ \hat{\psi}( x)\,   \hat{\psi}^{\dagger}(y)\big] \rangle    \\
    \langle  T\big[ \hat{\psi}^{\dagger}( x) \,  \hat{\psi}(y) \big]\rangle     &    \langle T\big[  \hat{\psi}^{\dagger}( x)\,   \hat{\psi}^{\dagger}(y) \big]\rangle     
\end{bmatrix}\, , \\
 iG_{ij}^{--} (x,y) &\equiv&\begin{bmatrix}
    \langle  \bar{T}\big[ \hat{\psi}( x)\,   \hat{\psi}(y) \big] \rangle    & \langle \bar{T}\big[  \hat{\psi}(x)\,   \hat{\psi}^{\dagger}(y)  \big]\rangle    \\
    \langle \bar{T}\big[  \hat{\psi}^{\dagger}( x) \,  \hat{\psi}(y) \big] \rangle     &    \langle \bar{T}\big[  \hat{\psi}^{\dagger}(x)\,   \hat{\psi}^{\dagger}(y)  \big]\rangle     
\end{bmatrix}\, ,  \\
 iG_{ij}^{-+} (x,y) &\equiv&\begin{bmatrix}
    \langle  \hat{\psi}(x)\,   \hat{\psi}(y) \rangle    & \langle  \hat{\psi}(x)\,   \hat{\psi}^{\dagger}(y) \rangle    \\
    \langle  \hat{\psi}^{\dagger}( x) \,  \hat{\psi}(y) \rangle     &    \langle  \hat{\psi}^{\dagger}(x)\,   \hat{\psi}^{\dagger}(y) \rangle     
\end{bmatrix}\, ,\\
 iG_{ij}^{+-} (x,y) &\equiv&\begin{bmatrix}
    \langle  \hat{\psi}(y)  \,   \hat{\psi}(x)\rangle    & \langle  \hat{\psi}^{\dagger}(y)\,   \hat{\psi}(x) \rangle    \\
    \langle   \hat{\psi}(y) \,  \hat{\psi}^{\dagger}( x) \rangle     &    \langle \hat{\psi}^{\dagger}(y) \,  \hat{\psi}^{\dagger}( x)  \rangle     
\end{bmatrix} \,,
\eea
where $T$ and $\bar{T}$ denote time ordering and anti-time ordering, respectively.
We will soon drop the squeezed state propagators to be consistent with \eqref{initCondNoSqueez}.
The divergent part of the derivative operator \eqref{operatorD} should be thought of part of the interaction term since it removes homogeneous contributions of the spatially non-local coupling.
\begin{figure}[!htb]
\minipage{0.48\textwidth}
  \includegraphics[width=\linewidth]{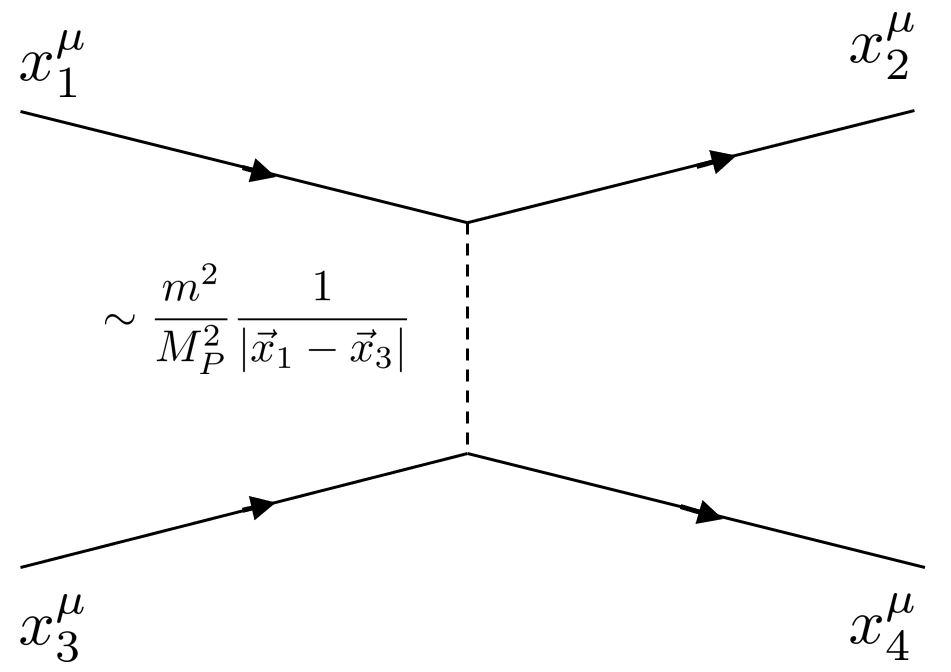}
  \caption{The Hartree vertex is local in time but non-local in space. The  separation between $\vec{x}_1$ and $\vec{x}_3$ (as well as between $\vec{x}_2$ and $\vec{x}_4$) is denoted by a dashed line.  }\label{fig:HartreeVert}
\endminipage \hfill
\minipage{0.48\textwidth}
  \includegraphics[width=\linewidth]{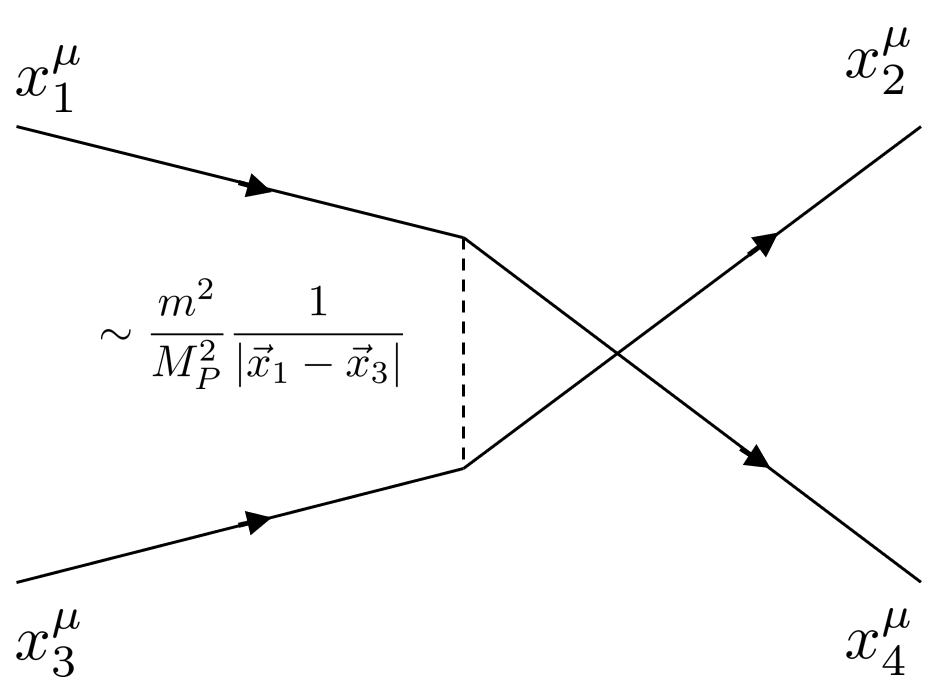}
  \caption{The exchange vertex is obtained from the Hartree vertex by exchanging the spatially separated coordinates $\vec{x}_2$ and $\vec{x}_4$.\\}\label{fig:ExchangeVert} 
\endminipage
\end{figure}
The two vertices $V_H$ and $V_E$ we use in \eqref{2piAction} are both symmetric under exchange of the first and last pair of coordinates and correspond to Hartree and exchange interaction (cf. figures ~\ref{fig:HartreeVert} and ~\ref{fig:ExchangeVert}), respectively, 
\bea
V_H^c (\eta_1,...,\eta_4,\vec{x}_1,...,\vec{x}_4)
      &\equiv&  i   \frac{3 c}{4 \pi  \beta \rho_0}   \frac{ \delta(\eta_1-\eta_2)\delta(\eta_1-\eta_3)\delta(\eta_1-\eta_4)}{ \|\vec{x}_1-\vec{x}_3\|}   \delta^3(\vec{x}_1-\vec{x}_2)  \delta^3(  \vec{x}_3-  \vec{x}_4)    \, ,\\
      V_E^c (\eta_1,...,\eta_4,\vec{x}_1,...,\vec{x}_4)
      &\equiv&   i \frac{3 c}{4 \pi  \beta\rho_0}    \frac{ \delta(\eta_1-\eta_2)\delta(\eta_1-\eta_3)\delta(\eta_1-\eta_4)}{ \|\vec{x}_1-\vec{x}_3\|}   \delta^3(\vec{x}_1-\vec{x}_4)   \delta^3(    \vec{x}_2-\vec{x}_3)   \,. \label{interactions}
\eea
Setting the variation of the 2PI effective action \eqref{2piAction} with respect to  $G_{ij}^{cd}$ to zero and multiplying the resulting equation again by $G_{ij}^{cd}$ , we obtain
\begin{multline}
 \begin{bmatrix}
   0  &  - i  \partial_{\eta} +{\beta \Delta_x}{\eta^{-2}}- {3}{\beta}^{-1} \big[\Delta^{-1}_x  (1)\big]  \\
    i \partial_{\eta}  +{\beta \Delta_x}{\eta^{-2}}  -  {3}{\beta}^{-1} \big[\Delta^{-1}_x  (1)\big] &   0   \end{bmatrix}^{ij} iG_{jk}^{cd}(\eta,\vec{x},\eta^{\prime},\vec{y})\\- i \frac{  c}{2\hbar} \int d^4z_1d^4z_2d^4z_3  V_{H+E}^c (x_1,z_3,z_1,z_2)\big[iG_{12}^{cc}(z_1,z_2)+iG_{21}^{cc}(z_1,z_2)\big] \begin{bmatrix}
   0 &  1      \\
 1     &      0  
\end{bmatrix}^{ij}   i G_{jk}^{cd}(z_3,\eta^{\prime},\vec{y})\\
- i \frac{ c }{2\hbar} \int d^4z_1d^4z_2d^4z_3  \begin{bmatrix}
    V_{H+E}^c (x_1,z_1,z_3,z_2)iG_{22}^{cc}(z_1,z_2)  &  0     \\
 0     &      V_{H+E}^c (z_1,x_1,z_2,z_3)iG_{11}^{cc}(z_1,z_2)   
\end{bmatrix}^{ij}  i G_{jk}^{cd}(z_3,\eta^{\prime},\vec{y}) \\\\=        i  c \delta^{cd}\delta^{i}_{\;k} \delta(\eta- \eta^{\prime}) \delta^3(\vec{x}-\vec{y})     \, .
\end{multline}
\begin{figure}[!htb]
  \includegraphics[width=\linewidth]{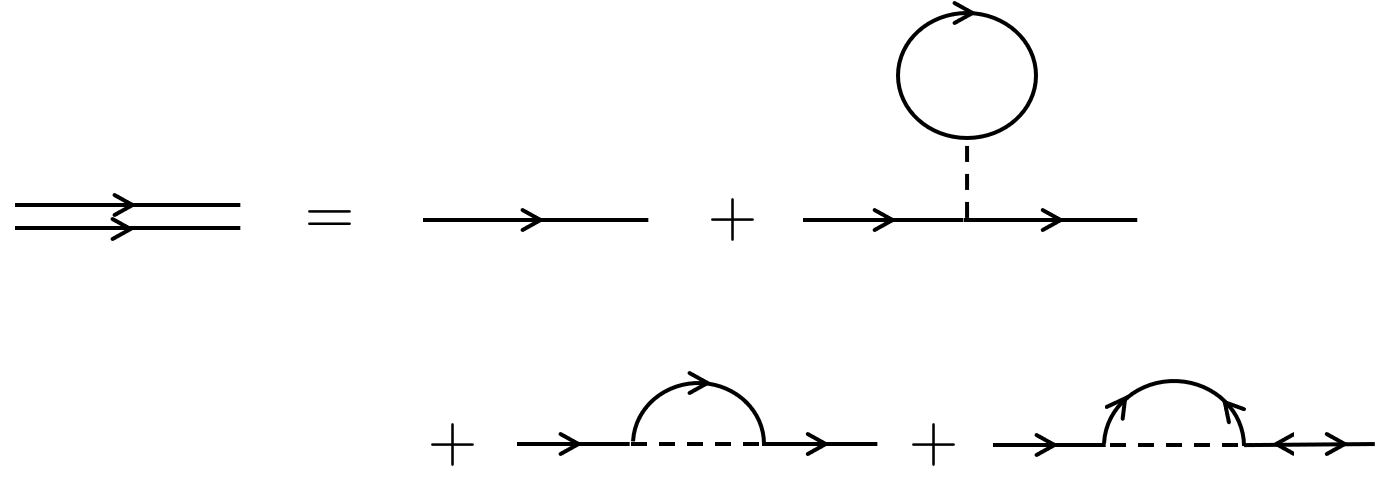}
  \caption{The 2PI equation for the full two-point function $G_{12}$ from the two-loop effective action \eqref{2piAction}. Dashed lines in the (spatial) loop denote spatial non-locality. Lines with two arrows denote the two-point functions $G_{11}$ and $G_{22}$ which are initially absent for non-squeezed states. For brevity we omitted three diagrams with identical topology but reversed flow in the loop.}\label{fig:2PIEQ}
\end{figure}
In the equations for $i G_{12}^{cd}$ and $i G_{21}^{cd}$ it is consistent within our approximation scheme \eqref{noSqueezenoVev} to drop the squeezing contributions $iG_{ii}^{cd}$. We then have the following equations for $iG_{21}^{-+}(\eta,\vec{x},\eta^{\prime},\vec{y})$ and $iG_{21}^{+-}(\eta^{\prime},\vec{y},\eta,\vec{x})$, which we will combine into a particle density,
\begin{multline}
 \big[  i \partial_{\eta}  -{(\eta)^{-2}} {\beta \Delta_x} \big]i G_{21}^{\mp \pm}(\eta,\vec{x},\eta^{\prime},\vec{y}) \\ +\frac{ 3}{8 \pi \beta \rho_0}  \int d^3z  \frac{1}{\|\vec{x}-\vec{z}\|}    \big[iG_{21}^{\mp \mp}(\eta, \vec{z},\eta,\vec{z})+iG_{12}^{\mp \mp}(\eta, \vec{z},\eta,\vec{z})-2\rho_0\big] iG_{21}^{\mp \pm}(\eta,\vec{x},\eta^{\prime} ,\vec{y})   
\\
+ \frac{ 3}{8 \pi \beta \rho_0} \int d^3z  \frac{1}{\|\vec{x}-\vec{z}\|}\big[iG_{21}^{\mp \mp}(\eta, \vec{z},\eta, \vec{x})+iG_{12}^{\mp \mp}(\eta, \vec{z},\eta, \vec{x})\big]i G_{21}^{\mp \pm}(\eta, \vec{z},\eta^{\prime} ,\vec{y}) \approx   0 \, .\label{G21EQ}
\end{multline}
\begin{multline}
 \big[  i \partial_{\eta^{\prime}}  +{(\eta^{\prime})^{-2}} {\beta \Delta_y} \big]i G_{21}^{\mp\pm }(\eta,\vec{x},\eta^{\prime},\vec{y}) \\ -\frac{ 3}{8 \pi \beta \rho_0}  \int d^3z  \frac{1}{\|\vec{y}-\vec{z}\|}    \big[iG_{21}^{\pm \pm}(\eta^{\prime}, \vec{z},\eta^{\prime},\vec{z})+iG_{12}^{\pm \pm}(\eta^{\prime}, \vec{z},\eta^{\prime},\vec{z})-2\rho_0\big] iG_{21}^{\mp\pm}(\eta,\vec{x},\eta^{\prime} ,\vec{y})   
\\
-\frac{ 3}{8 \pi \beta \rho_0} \int d^3z  \frac{1}{\|\vec{y}-\vec{z}\|}\big[iG_{21}^{\pm \pm}(\eta, \vec{y} ,\eta, \vec{z})+iG_{12}^{\pm \pm}(\eta,\vec{y},\eta, \vec{z} )\big]i G_{21}^{\mp\pm}(\eta,\vec{x} ,\eta^{\prime} ,\vec{z})   \approx     0 \, ,\label{G21EQalt}
\end{multline}
 We will make use of the following statistical (Hadamard) two-point function
\bea
F (\eta,\vec{x},\eta^{\prime},\vec{y})\equiv F_{21} (\eta,\vec{x},\eta^{\prime},\vec{y}) \nonumber &=& \frac{1}{2}\big[ iG_{21}^{-+} (\eta,\vec{x},\eta^{\prime},\vec{y})+iG_{21}^{+-} (\eta,\vec{x},\eta^{\prime},\vec{y}) \big]\\
&=& \frac{1}{2}\big\langle \big\lbrace  \hat{\psi}^{\dagger}(\eta, \vec{x})\, ,   \hat{\psi}(\eta^{\prime},\vec{y}) \big\rbrace \big\rangle\, .
\eea
The spectral density
\be
i \rho^s_{21} (\eta,\vec{x},\eta^{\prime},\vec{y}) =\big[i G_{21}^{-+} (\eta,\vec{x},\eta^{\prime},\vec{y})-iG_{21}^{+-} (\eta^{\prime},\vec{x},\eta,\vec{y}) \big]= i \big\langle \big[  \hat{\psi}^{\dagger}(\eta, \vec{x})\, ,   \hat{\psi}(\eta^{\prime},\vec{y}) \big] \big\rangle\, ,
\ee
will drop out once we evaluate the coincident time limit.
We use collective (average) and difference coordinates to define
\be
F (\eta,\vec{X},\vec{r}) \equiv F (\eta,\eta^{\prime} = \eta,\vec{x}=\vec{X}+\vec{r}/{2},\vec{y}=\vec{X}-\vec{r}/{2}) \, .
\ee
Adding up the equations for $G_{21}^{\pm \mp}$ in \eqref{G21EQ} and \eqref{G21EQalt} we find in the coincident time limit,
\begin{multline}
 \Big[  i \partial_{\eta} +2{\beta }{\eta^{-2}} {\frac{\partial}{\partial \vec{X}} \cdot \frac{\partial}{\partial \vec{r}} }  \Big] F(\eta,\vec{X},\vec{r}) 
+\frac{ 3}{4 \pi \beta \rho_0} \int\frac{ d^3z  }{z}\Big[F(\eta, \vec{z} + \vec{X}+\vec{r}/2,0) - F(\eta, \vec{z}+ \vec{X}-\vec{r}/2 ,0)\Big]F(\eta,\vec{X} ,\vec{r})  \\
+  \frac{ 3}{8 \pi \beta \rho_0}\int\frac{d^3z}{z}\Big[ F(\eta, \vec{X}+(\vec{r}+ \vec{z} )/2, \vec{z})+ F(\eta, \vec{X}+(\vec{r}+ \vec{z} )/2, -\vec{z}) \Big]F(\eta, \vec{X}+\vec{z}/2 ,\vec{r}+ \vec{z}  )  \\
-  \frac{ 3}{8 \pi \beta\rho_0}\int\frac{d^3z}{z}\Big[ F(\eta, \vec{X}-(\vec{r}-\vec{z})/2 , \vec{z}   )+F(\eta, \vec{X}-(\vec{r}-\vec{z})/2 ,- \vec{z}   ) \Big]F(\eta, \vec{X}+\vec{z}/2 , \vec{r}-\vec{z} ) 
 =0    \, . \label{collectAndDiffCoord}
\end{multline}
The two-loop effective action \eqref{2piAction} contains only quartic interactions such that the resulting scalar self-mass in equation \eqref{collectAndDiffCoord} contains no dissipative contributions (the imaginary part of the self-mass vanishes) which is why the equations close for equalt-time two-point functions.
We see that the homogeneous and isotropic equation is solved by a function $F_{\text{hom}}( r)$ which is constant in time and constant in the collective coordinate $\vec{X}$,
\be
F_{\text{hom}}(\eta, \vec{X},\vec{r}) =F_{\text{hom}}( r) \quad \text{with} \quad F_{\text{hom}}( 0)=\rho_0\, ,
\ee
which matches the initial conditions at spatial infinity \eqref{initCondNoSqueez}.
Let us switch to momentum space and introduce the inhomogeneous Wigner transformation,
\be
F(\eta,\vec{k},\vec{p}) = \frac{1}{(2 \pi \hbar)^6} \int d^3 X e^{- \frac{i}{\hbar} \vec{k}\cdot \vec{X}}\int d^3 r e^{- \frac{i}{\hbar} \vec{p}\cdot \vec{r}} F\big(\eta,  \vec{X}, \vec{r}\big)\, .
\ee
We emphasize that were are counting both momenta, small scale momentum $\vec{p}$ and large scale momentum $\vec{k}$, in units of energy. 
We then have
\begin{multline}
 \big[  i \partial_{\eta} -2 {(\hbar\eta)^{-2}{\beta }} \vec{k} \cdot \vec{p} \big] F (\eta,\vec{k},\vec{p}) 
+    \frac{ 3 \hbar^2 }{2 \beta \rho_0}  \int d^3 w \int d^3 u F(\eta,\vec{w}, \vec{u})\\\times\Bigg[    \| \vec{p}+\vec{u}+(\vec{k}-\vec{w}) /2)\|^{-2}    F(\eta,\vec{k}-\vec{w},  \vec{p}-\vec{w} /2) 
-  \| \vec{p}+\vec{u}-(\vec{k}-\vec{w}) /2)\|^{-2}   F(\eta,\vec{k}-\vec{w},  \vec{p}+\vec{w} /2)\\
+\| \vec{p}-\vec{u}+(\vec{k}-\vec{w}) /2)\|^{-2}    F(\eta,\vec{k}-\vec{w},  \vec{p}-\vec{w} /2) 
-  \| \vec{p}-\vec{u}-(\vec{k}-\vec{w}) /2)\|^{-2}   F(\eta,\vec{k}-\vec{w},  \vec{p}+\vec{w} /2)\\
+  2 w^{-2} \Big( F(\eta,\vec{k}-\vec{w}, \vec{p}-\vec{w} /2) -  F(\eta,\vec{k}-\vec{w}, \vec{p}+\vec{w} /2) \Big)\Bigg] =0 \, .
\end{multline}
Our next goal is to expand around a homogeneous Maxwellian distribution and see which differences we get (at least in the linear theory) in comparison to classical particle cold dark matter. It will turn out to be convenient if we
rescale all momenta  and times
\be
\vec{p} \rightarrow  \vec{p} \alpha^{1/2}\, , \quad \vec{k} \rightarrow   \vec{k} \alpha^{1/2}\, , \quad \alpha \equiv m  k_B T\, , \quad \eta \longrightarrow \eta_I \tau   = \frac{\tau}{2 \mathcal{H}_I}\, ,  \label{rescalings}
\ee
such that the quantities on the right-hand-side of \eqref{rescalings} are dimensionless. The dimensionless time $\tau$ is nothing but the square-root of the scale factor $a$. The parameter  $\alpha$ is the geometric mean between the particles mass $m$ and temperature parameter $k_B T$ with $k_B$ being the Boltzmann constant. Thus, the parameter  $\alpha$ corresponds to the averaged particle moment $\langle p^2 \rangle$ where the expectation value denotes here the integral against a particle distribution in momentum space which we choose to be a Maxwellian distribution. 
Moreover, it will be handy to define the parameter
\be
\xi \equiv \frac{ \alpha \beta}{\hbar ^2 \eta_I} = \frac{m k_B T}{\hbar^2 \eta_I}  \frac{\hbar \eta_I^2}{2m} =  \frac{ k_B T }{ \hbar \mathcal{H}_I } \, .
\ee
We will see that the parameter $\xi$ will decide on which time-scales the exchange interaction term can become important if we are working on scales $k \ll p$.
Moreover, we rescale the coincident Hadamard function as
\be
F \longrightarrow \alpha^{-3} \rho_0 F\, , 
\ee
so that the $p$-integral over its inhomogeneous part yields the density contrast.
We also assume further, that is only a function of the moduli $k$ and $p$ as well as its scalar product 
\be
F(\tau,\vec{k},\vec{p} ) = F(\tau, k ,p ,\mu )\, , \quad \mu = \frac{\vec{p}\cdot \vec{k}}{p k }\, ,
\ee
and expand it as\footnote{We note that the perturbations $\delta F$ should in principle be multiplied by stochastic variables $\hat{a}_{\vec{k}}$ such that the perturbations of the two-point functions $F(\eta,\vec{k},\vec{p} )$ are stochastic variables in a cosmological context.}
\be
F(\tau, k ,p, \mu) = (2 \pi )^{-3/2} \delta^3(\vec{k}) e^{-p^2/2} + \delta F (\tau,k , p ,\mu)\,.
\ee
We have
\begin{multline}
 \Big[  i \partial_{\tau} -2 \tau^{-2}\xi k p\mu  \Big]\delta F (\tau,k,p,\mu)  \\ + \frac{6}{\xi k^2}  (2 \pi )^{-3/2}\exp \Big[-\frac{p^2}{2} - \frac{k^2}{8} \Big] \, \text{sinh}\Big[  \frac{p k \mu}{2}\Big] \int d^3 u\Bigg[1+ \frac{k^2}{2\|\vec{p}+\vec{u}\|^{2}  }+ \frac{k^2}{2\|\vec{p}-\vec{u}\|^{2}  }\Bigg]\delta F(\tau,k,u, \mu_{k,u})\\
+\frac{6}{\xi }2^{1/2}
 \delta F (\tau,k,p,\mu)
   \Bigg\lbrace
   \frac{\text{DawsonF}\Big[ 2^{-1/2} \|\vec{p}- \vec{k}/2\|\Big]}{\|\vec{p}- \vec{k}/2\|}
    -
     \frac{\text{DawsonF}\Big[ 2^{-1/2} \|\vec{p}+ \vec{k}/2\|\Big]}{\|\vec{p}+ \vec{k}/2\|}\Bigg\rbrace
\\+ \frac{3}{2 \xi }\int d^3 w \int d^3 u \delta F(\tau,w,u, \mu_{w,u})\\\times\Bigg[    \| \vec{p}+\vec{u}+(\vec{k}-\vec{w}) /2)\|^{-2}    \delta F(\tau,\vec{k}-\vec{w},  \vec{p}-\vec{w} /2) 
-  \| \vec{p}+\vec{u}-(\vec{k}-\vec{w}) /2)\|^{-2}   \delta F(\tau,\vec{k}-\vec{w},  \vec{p}+\vec{w} /2)\\
+\| \vec{p}-\vec{u}+(\vec{k}-\vec{w}) /2)\|^{-2}    \delta F(\tau,\vec{k}-\vec{w},  \vec{p}-\vec{w} /2) 
-  \| \vec{p}-\vec{u}-(\vec{k}-\vec{w}) /2)\|^{-2}   \delta F(\tau,\vec{k}-\vec{w},  \vec{p}+\vec{w} /2)\\
+  2 w^{-2} \Big( \delta F(\tau,\vec{k}-\vec{w}, \vec{p}-\vec{w} /2) - \delta  F(\tau,\vec{k}-\vec{w}, \vec{p}+\vec{w} /2) \Big)\Bigg]=0\, , \label{diffEqF}
\end{multline}
where we made use of the Dawson integral
\be
\text{DawsonF}(z) = e^{-z^2}\int_0^z e^{y^2}dy =  z \, e^{-z^2} {{_1}F_1} \Big( \frac{1}{2}\,, \frac{3}{2}\,, z^2\Big) \,,
\ee
where ${{_1}F_1}$ is the confluent hypergeometric function of the first kind.
Let us define
\be
F(\tau,k) \equiv \int d^3 u F(\tau,k,u, \mu_{k,u})\, ,
\ee
and contrast equation \eqref{diffEqF} with the perturbed Vlasov description in the truncated equation \eqref{Vlasov}. 
We realize that the terms
\begin{multline}
\mathcal{V}[\delta F] \equiv \Big[i \partial_{\tau} -2 \tau^{-2}\xi k p\mu  \Big]\delta F (\tau,k,p,\mu)
+ \frac{6}{\xi k^2}  (2 \pi )^{-3/2}\exp \Big[-\frac{p^2}{2} - \frac{k^2}{8} \Big] \, \text{sinh}\Big[  \frac{p k \mu}{2}\Big] \delta F(\tau,k)\\
+ \frac{3}{ \xi }\int d^3 w  \delta F(\tau,w)   w^{-2} \Big( \delta F(\tau,\vec{k}-\vec{w}, \vec{p}-\vec{w} /2) - \delta  F(\tau,\vec{k}-\vec{w}, \vec{p}+\vec{w} /2) \Big) \,, \label{closetoVlasov}
\end{multline}
should correspond to the full non-linear Vlasov equation if we work in the limit where particle momenta are much bigger than large-scale momenta ($ p \sim 1 \gg k$) which is amply satisfied for a cold dark matter scenario with galactic scales around $\sim \text{Mpc} \gg \alpha^{-1/2}$. There are, however,  differences and we first note the appearance of a "$\text{sinh}$" in place of the partial derivative $\partial_{\vec{p}}$ acting on the background phase-space density. As we will see, the "$\text{sinh}$" term yields the same results for the linear theory on galactic scales if other terms can be neglected. The second difference is the non-linear term in \eqref{closetoVlasov} which, however, may be converted into a partial derivative  for $k/p \ll 1$  as it appears in the Vlasov equation.
In addition to the Vlasov-like terms in \eqref{diffEqF}, we note the appearance of exchange interaction corrections  which are of order $\sim k^2/p^2$,
\begin{multline}
\mathcal{E}[\delta F] \equiv  \frac{3}{\xi k^2}  (2 \pi )^{-3/2}\exp \Big[-\frac{p^2}{2} - \frac{k^2}{8} \Big] \, \text{sinh}\Big[  \frac{p k \mu}{2}\Big] \int d^3 u\Bigg[ \frac{k^2}{\|\vec{p}+\vec{u}\|^{2}  }+ \frac{k^2}{\|\vec{p}-\vec{u}\|^{2}  }\Bigg]\delta F(\tau,k,u, \mu_{k,u})
\\+ \frac{3}{2 \xi }\int d^3 w \int d^3 u \delta F(\tau,w,u, \mu_{w,u})\\\times\Bigg[    \| \vec{p}+\vec{u}+(\vec{k}-\vec{w}) /2)\|^{-2}    \delta F(\tau,\vec{k}-\vec{w},  \vec{p}-\vec{w} /2) 
-  \| \vec{p}+\vec{u}-(\vec{k}-\vec{w}) /2)\|^{-2}   \delta F(\tau,\vec{k}-\vec{w},  \vec{p}+\vec{w} /2)\\
+\| \vec{p}-\vec{u}+(\vec{k}-\vec{w}) /2)\|^{-2}    \delta F(\tau,\vec{k}-\vec{w},  \vec{p}-\vec{w} /2) 
-  \| \vec{p}-\vec{u}-(\vec{k}-\vec{w}) /2)\|^{-2}   \delta F(\tau,\vec{k}-\vec{w},  \vec{p}+\vec{w} /2)\Bigg] \,. \label{exchangeCorrVlas}
\end{multline}
Since, we are for the moment interested in scales larger or at most comparable to galactic scales, we will assume from now on the limit $k \ll 1$ and postpone the study of this type of corrections for future research (however, we expect small scale effects similar to ones for fuzzy dark matter as described for example in \cite{Hui:2016ltb}). 
There is another  term originating from linearly expanding the exchange interaction term \eqref{interactions},
\begin{multline}
\mathcal{F}[\delta F] \equiv \frac{6}{\xi }2^{1/2}
 \delta F (\tau,k,p,\mu)
   \Bigg\lbrace
   \frac{\text{DawsonF}\Big[ 2^{-1/2} \|\vec{p}- \vec{k}/2\|\Big]}{\|\vec{p}- \vec{k}/2\|}
    -
     \frac{\text{DawsonF}\Big[ 2^{-1/2} \|\vec{p}+ \vec{k}/2\|\Big]}{\|\vec{p}+ \vec{k}/2\|}\Bigg\rbrace \,. \label{exchangeCorrVlas2}
\end{multline} As we will discuss shortly, it gives rise to late-time corrections and is {\it{not}} $k^2/p^2$ suppressed in contrast to all other terms originating from the exchange interaction.

\begin{figure}
\begin{center}
  \includegraphics[width=0.65\textwidth]{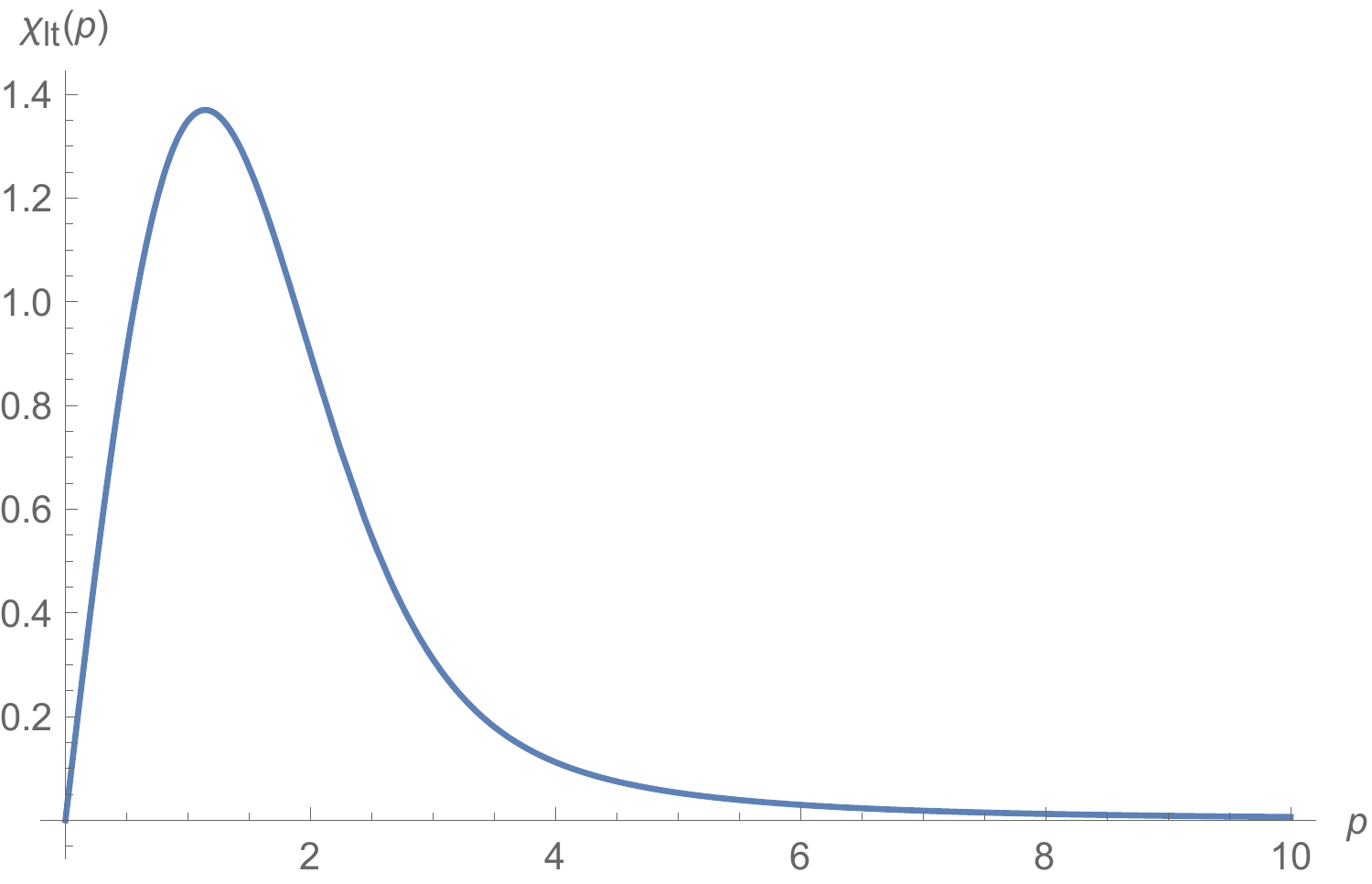}
  \caption{The function  $\chi_{\text{lt}} $ defined in \eqref{dawsonType} dominating the late-time behaviour of the phase factor \eqref{phasefactor} in the $k \ll 1$ expansion.}\label{fig:dawson}
    \end{center}
\end{figure}

We would now like to proceed studying \eqref{diffEqF}, however, without taking moments in $p$ to avoid arguing about the smallness of higher moments. Therefore, it is convenient to convert \eqref{diffEqF} into an integral equation for the density contrast by defining
\be
\chi (\tau, k ,p , \mu) \equiv  2\tau^{-1}\xi k p\mu + \tau \frac{6}{\xi }2^{1/2}
   \Bigg\lbrace
   \frac{\text{DawsonF}\Big[ 2^{-1/2} \|\vec{p}- \vec{k}/2\|\Big]}{\|\vec{p}- \vec{k}/2\|}
    -
     \frac{\text{DawsonF}\Big[ 2^{-1/2} \|\vec{p}+ \vec{k}/2\|\Big]}{\|\vec{p}+ \vec{k}/2\|}\Bigg\rbrace\, , \label{phasefactor}
\ee
with the series expansion in $k \ll 1 \sim p $,
\be
\chi (\tau, k ,p , \mu) =  2\tau^{-1}\xi k p\mu +   \frac{6 \mu k \tau}{\xi p^2 }
   \Big[
  2^{1/2}(1+ p^2) {\text{DawsonF}\Big[ 2^{-1/2} p \Big]} 
  -p 
   \Big] +\mathcal{O}\big( k^3 \big)\, . \label{phaseFactorExp}
\ee
We note that the p-dependent factor in the expansion of the late-time term \eqref{phaseFactorExp},
\be
\chi_{\text{lt}} (p)\equiv  \frac{3}{ p^2 }
   \Big[
  2^{1/2}(1+ p^2) {\text{DawsonF}\Big[ 2^{-1/2} p \Big]} 
  -p 
   \Big]\, ,\label{dawsonType}
\ee 
is of order $1$ for $p \sim 1$  (cf. figure \ref{fig:dawson})  and thus, phase corrections  due to the exchange interaction term become only important at late times if we work in the limit $k \ll 1$. 
The transition time from which on the late time phase factor dominates is given by
\be
\eta_{\text{trans}} \equiv \xi \eta_I  = \frac{k_B T}{\hbar \mathcal{H}_I} \eta_I   \, , \label{transtime}
\ee
which is a very large time even on cosmological scales unless the particle temperature is extremely small.
We now make use of the phase definition \eqref{phasefactor} and integrate equation \eqref{diffEqF} in time. As just discussed below \eqref{diffEqF}, we neglect the $p^2/k^2$ corrections due to the exchange interaction terms and are left with
\begin{multline}
 \delta F (\tau,k) \approx \int d^3 p\,  \text{exp}\Big[ i \chi(\tau,k,p,\mu)-i\chi(\tau_I,k,p,\mu) \Big]  \delta F_I (k,p,\mu) \\  +i\int d^3 p\,  \text{exp}\Big[i \chi(\tau, k ,p , \mu)\Big]\int^{\tau}_{1} d \bar{\tau} 
\text{exp}\Big[-i \chi(\bar{\tau}, k ,p , \mu)\Big] \Bigg\lbrace \frac{6}{\xi k^2}  (2 \pi )^{-3/2} \exp \Big[-\frac{p^2}{2} - \frac{k^2}{8} \Big] \, \text{sinh}\Big[  \frac{p k \mu}{2}\Big] \delta F(\bar{\tau},k)
\\-  \frac{3}{\xi }\int d^3 w \int d^3 u \frac{ \delta F(\bar{\tau},w,u, \mu_{w,u})}{w^2}     \Big( \delta F(\bar{\tau},\vec{k}-\vec{w}, \vec{p}-\vec{w} /2) - \delta  F(\bar{\tau},\vec{k}-\vec{w}, \vec{p}+\vec{w} /2) \Big)\Bigg\rbrace  \, . \label{intgegralEQ}
\end{multline}
\paragraph{Case $ \xi^2 \gg  a(\eta)$, Hartree interaction phase   dominates.}
First, we study the integral equation \eqref{intgegralEQ} for   dimensionless times $\tau$ which are much smaller then the parameter $\xi$ (despite this, they can still correspond to galactic time scales $\eta_{\text{final}} \sim 10^{5-10} \eta_I$),
\be
\xi = \frac{k_B T}{\hbar \mathcal{H}_I} \gg \tau = \sqrt{a(\eta)}\, .
\ee
We can then write equation \eqref{intgegralEQ} as
\begin{multline}
 \delta F (\tau,k) \approx \int d^3 p\,  \text{exp}\big[ 2i (\tau^{-1}-1)\xi k p\mu \big] \delta F_I (k,p,\mu) \\ +\frac{6}{k^2 \xi}\int^{\tau}_{1} d \bar{\tau}\,   \text{exp}\Big[ - \frac{2k^2 \xi^2 (\tau - \bar{\tau})^2}{\tau^2 \bar{\tau}^2}\Big] \, \text{sin}\Big[  \frac{k^2 \xi (\tau - \bar{\tau})}{ \tau \bar{\tau}}\Big] \delta F(\bar{\tau},k) 
\\
- \frac{6}{\xi } \int d^3 p\,  \int^{\tau}_{1} d \bar{\tau} 
\text{exp}\big[2i (\tau^{-1}- \bar{\tau}^{-1})\xi k p\mu \big] \int d^3 w \frac{\delta F(\bar{\tau},w)}{w^2}   \text{sin}\Big[  \frac{k  w \mu_{kw}\xi (\tau - \bar{\tau})}{ \tau \bar{\tau}}\Big]  \delta F(\bar{\tau},\vec{k}-\vec{w}, \vec{p})   \, . \label{intNonlinNoExPot}
\end{multline}
We discover two scales in expression \eqref{intNonlinNoExPot}.
The first scale appears in the oscillatory terms,
\be
{k_{\text{osc}}(\eta)} \equiv   \sqrt{ m a(\eta) \times \hbar a(\eta) H(\eta)}= \sqrt{ m a(\eta) k_{H}(\eta)} = a(\eta)^{1/4} \frac{\alpha^{1/2}}{\xi^{1/2}} = \alpha^{1/2} \Big(\frac{\tau}{\xi}   \Big)^{1/2}\, , \label{kOSC}
\ee
where we introduced the Hubble scale
\be
k_H(\eta) \equiv \hbar a(\eta) H(\eta) = \hbar \mathcal{H}(\eta)\,.
\ee
The scale $k_{\text{osc}}$ in \eqref{kOSC} is the geometric mean between the scale of relativistic effects and the sub-Hubble scale
\be
k_{\text{rel}} \gtrsim  k_{\text{osc}}  \gtrsim k_{H} \, ,
\ee 
and we suspect that structure formation is inhibited at these scales due to oscillatory solutions.
The second important scale in expression \eqref{intNonlinNoExPot} appears in the exponential for the linear term.
The question, whether this exponential is important may be answered by referring to the scale
\be
k_{\xi}(\eta) \equiv \frac{ \alpha^{1/2}  \tau }{\xi} = k_H(\eta) a \Big(\frac{m}{k_B T}   \Big)^{1/2} =  k_{\text{osc}}(\eta)  \Big(\frac{ k_H(\eta) a}{k_B T}   \Big)^{1/2}=   k_{\text{osc}}(\eta) \frac{a(\eta)^{1/4}}{\xi^{1/2}}      =   k_{\text{osc}}(\eta) \Big(\frac{\tau}{\xi}   \Big)^{1/2}   \, . \label{kXi}
\ee
Relative to sub-Hubble scales, the scale $k_{ \xi}$ is in reach for light and warm particles.
Since we are working  in the limit $\xi \gg \tau$ in this paragraph, we have
\be
k_{\text{osc}} \gg k_{\xi} \,,
\ee
such that the exponential suppression in the linear term in  \eqref{intNonlinNoExPot} begins before oscillatory contributions  become important.
It is of course tempting to study the full $k$-dependence in the linearized version of equation \eqref{intNonlinNoExPot}. However, we are not aware of a solution in terms of the exponential and sinusoidal kernel
\be
 K[k,\tau, \bar{\tau}] \equiv \text{exp}\Big[ - \frac{2k^2 \xi^2 (\tau - \bar{\tau})^2}{\tau^2 \bar{\tau}^2}\Big] \, \text{sin}\Big[  \frac{k^2 \xi (\tau - \bar{\tau})}{ \tau \bar{\tau}}\Big]\, , \label{kernel}
\ee
and leave it for future research.
For cold dark matter it is now a reasonable scenario to assume\footnote{For the cold dark matter paradigm we have the limit $m/(k_B T) \gtrsim 10^{12}$ where WIMPs are far away from this limit with $m/(k_B T) \gtrsim 10^{24}$ \cite{Armendariz-Picon:2013jej}.}
\be
k  \ll k_{\xi} (\eta_I)\ll k_{\xi} (\eta)\, ,
\ee
in which case
\be
\int d^3 p\,  \text{exp}\Big[ 2i (\tau^{-1}-1)\xi k p\mu \Big] \delta F_I (k,p,\mu) \longrightarrow  \delta F_I (k)\quad \text{for} \quad \tau \gg 1\,, \quad k  \ll k_{\xi}\, .
\ee
Moreover, the exponential suppression in \eqref{intNonlinNoExPot} is negligible in this scenario
\be
 \text{exp}\Big[ - \frac{2k^2 \xi^2 (\tau - \bar{\tau})^2}{\tau^2 \bar{\tau}^2}\Big] \longrightarrow 1 \quad \text{for} \quad   k \ll k_{\xi}\,.
\ee
Since we are working out the case $\xi \gg \tau$ in this paragraph, the sine can also be expanded around zero.
We then have
\be 
 \delta F_{\text{lin}} (\tau,k) \approx \delta F_I ( k)  +6 \int^{\tau}_{1} d \bar{\tau}\,     \frac{(\tau - \bar{\tau})}{ \tau \bar{\tau}} \delta F_{\text{lin}} (\bar{\tau},k)  \, ,
\ee
which is solved at late times by the standard linear cold dark matter evolution
\be
\delta F_{\text{lin}} (\eta,k) {\longrightarrow} \frac{3}{5}a(\eta)\delta F_I (k)\, , \quad \text{for} \quad k \ll k_{\text{osc}}(\eta)\, . \label{linCDM}
\ee
Although the form \eqref{intNonlinNoExPot} differs slightly from the Vlasov description \eqref{Vlasov}, the study of non-linear evolution is still highly non-trivial and we leave the discussion of approximations and perturbative expansions for the future.
Let us now discuss the other limit that brings the exchange interaction term into play.
\paragraph{Case $\xi^2 \ll a(\eta)$, exchange interaction phase dominates.} 
For this case, we approximate the phase-factor by \eqref{phaseFactorExp} and drop the free-streaming contributions $\sim \tau^{-1}$. We will be able to say something about the linear evolution. Unfortunately, we are not in the position to perform the full momentum integral for the linear term as we could in the case $\xi^2 \gg  a(\eta)$, which is why we have to restrict ourselves to
\be
k \ll k_{\xi}\, .
\ee
The linearized integral equation \eqref{intgegralEQ} then reads
\be
 \delta F_{\text{lin}} (\tau,k) \approx \delta F_I (\tau,k) - \frac{3}{\xi^2}\int^{\tau}_{1} d \bar{\tau} (\tau-\bar{\tau}) \delta F_{\text{lin}}(\bar{\tau},k)
  \, ,
\ee
and is quickly solved in terms of the scale factor by
\be
 \delta F_{\text{lin}} (\eta,k) = \delta F_I (k) \cos \Bigg[ \frac{\sqrt{3}}{\xi} (\sqrt{a(\eta)}-1 )\Bigg] \stackrel{a \gg a_I}{\longrightarrow}\delta F_I (k) \cos \Bigg[ \frac{\sqrt{3}a(\eta)}{\xi} \Bigg]\, .
\ee
We conclude that there is no linear growth for a small enough parameter $\xi$ such that at late times $\xi \ll \tau$ (on scales $k \ll k_{\text{osc}}$).
Thus, the effect of the exchange interaction term is to hinder the growth of linear perturbations for large distances at late time where late times are defined to be greater than the transition $\eta_{\text{trans}}$ given in \eqref{linCDM} which depends on the temperature of cold dark matter. If we demand as a rough estimate that the observed power spectrum for linear modes does not oscillate around a constant value, field-theoretic corrections yield a lower bound on the temperature of cold dark matter,
\be
k_B T \gtrsim \frac{a_{\text{today}}}{a_{\text{I}}} \frac{H_0}{\hbar} \approx 10^{-38} \text{GeV}\, ,
\ee 
where $a_I$ is the scale factor at the beginning of the matter dominated epoch and $H_0$ the Hubble rate today.
\pagebreak
\section{Conclusion and outlook}
We present a new formalism for deriving the non-relativistic limits starting with a covariant QFT tree-level action in which a real scalar field couples minimally to gravity. The key ingredients are to introduce an approximate diagonal field representation   \eqref{defPsi} for cosmological space-times and integrate out the gravitational constraint fields in a perturbative expansion. We focus on scalar perturbations in the longitudinal gauge but the formalism can be straightforwardly adapted to include also  tensor and vector gravitational perturbations and even modified gravitational theories to study their non-relativistic limits in a controlled way. We derive a general non-relativistic, non-local action \eqref{finalPsiGenState} for gravitational interacting matter on sub-Hubble scales that makes no reference to a particular state in the sense that it can contain a condensate, as well as squeezed contributions (all correlators in \eqref{allcorrs}). 

Let us summarize the assumptions and approximations which are needed to arrive at the final action \eqref{finalPsiGenState}. First of all, we neglect vector and tensor perturbations in the metric and linearize around a homogeneous, spatially flat FRLW-metric with scalar perturbations in the longitudinal gauge with gravitational potentials $\Phi_G$ and $\Psi_G$,
\be
\mathcal{O} \big( \Phi_G\, , \Psi_G \big) = \varepsilon_g^2 \ll 1\, .
\ee
Secondly, by expanding around these potentials, we assume that gravitational boundary terms and zero-mode fluctuations around the classical and a priori free-to-choose FRLW-metric to be negligible. However, for a consistent perturbative expansion of the action we ultimately pick the classical FRLW-metric in such a way that the boundary conditions \eqref{delteE0} and \eqref{delteE1}, which are nothing but the homogeneous semi-classical Einstein equations, are satisfied. Thirdly, we are working in a non-relativistic limit with
\be
\mathcal{O} \Big(  \frac{\hbar\|\nabla\|}{m} \Big) = \varepsilon_{\text{nr}} \ll 1\, .
\ee 
Spatial derivatives $\nabla = \nabla_{\vec{x}}$ acting on matter fields $\psi(\vec{x})$ will be mapped on particle momenta $\vec{p}$ and long-distance gradients $\nabla_{\vec{X}}\sim \hbar \vec{k}$ once two-point functions of fields such as $\langle {\psi}^{\dagger} (\eta,\vec{x})  {\psi}(\eta,\vec{y}))\rangle$ are mapped to a particle phase-space density $f(\eta, \vec{p},\vec{X})$. Thus, assuming $\hbar \|\nabla\| \ll m $ corresponds to assuming physical momenta $p$ and inverse distance scales $L^{-1}\sim k$ of the underlying physical problem to be much smaller than the scale set by the mass $m$.
Moreover, we consider the case where the mass $m$ is much bigger than the Hubble rate or its logarithmic derivative 
\be
\mathcal{O} \Big(   \frac{\hbar\mathcal{H}}{m a}\, , \frac{\hbar\mathcal{H}^{\prime}}{\mathcal{H} m a }\Big) = \varepsilon_{\text{\scriptsize H/m}} \ll 1\, .
\ee
Finally, we focus on the sub-Hubble limit relevant for structure formation and introduce the perturbation parameter
\be
\mathcal{O}\Bigg(\, \frac{\mathcal{H}^2}{\|\Delta\| }\, , \frac{\mathcal{H}^{\prime}}{\|\Delta\| } \,\Bigg) = { \varepsilon_{\text{\scriptsize H/k}}} \ll 1\,.
\ee

For the scope of this paper we study the derived action \eqref{finalPsiGenState} for a non-squeezed state without condensate contributions and derive the corresponding 2PI two-loop effective action. Because this two-loop action contains only quartic interactions it is non-dissipative, which allows us to get closure for the dynamics of the coincident two-point functions. The resulting equations have a form of classical kinetic equations. By performing an inhomogeneous Wigner transformation, we derive the dynamics of the dark matter phase space density \eqref{diffEqF} and compare it to the standard Vlasov equation describing particle cold dark matter. For large galactic scales and masses, we recover a description close to particle cold dark matter which is confirmed by the linear evolution \eqref{linCDM}. This is, however, the case only if the particles temperature is much bigger than the Hubble scale, since otherwise the exchange interaction (absent in the Vlasov description) becomes important at late times. 
Another important result of this work is that we identify two scales at which we suspect density perturbations to deviate significantly from the standard CDM evolution. These are the scale $k_{\text{osc}}$ \eqref{kOSC} between the relativistic and the sub-Hubble scale  and the scale $k_{\xi}$ \eqref{kXi} related to the ratio between dark matter temperature and its mass.  
These results were derived in the limit where particle momenta $p$ are much bigger than the large scale momentum $k$ (or in other words where the distances of the system under study are much bigger than de Broglie wavelength). However, the general formula \eqref{diffEqF} can  be used to study also the case $k \sim p$ where we expect new effects due to the exchange interaction term \eqref{interactions} to kick in. 

Another route of investigation is  to start from the more general non-relativistic action \eqref{finalPsiGenState} we derive in section \ref{genStateSecACtionc} and to study the interplay between different state contribution, i.e. the influence of particle dark matter on fuzzy dark matter and vice versa.

\paragraph{Acknowledgments.}
This work is part of the research programme of the Foundation for Fundamental Research on Matter (FOM), which is part of the Netherlands Organisation for Scientific Research (NWO). This work is in part supported by the D-ITP consortium, a program of the Netherlands Organization for Scientific Research (NWO) that is funded by the Dutch Ministry of Education, Culture and Science (OCW). 
\pagebreak
\bibliographystyle{abbrv}
\bibliography{Biblio}
\end{document}